\RequirePackage{fix-cm}
\documentclass[smallextended]{svjour3}       
\smartqed  
\usepackage{graphicx}

\usepackage{amssymb}
\usepackage{amsmath}
\usepackage{bm}
\usepackage{mathptmx}

\begin{document}

\title{Physical quantities and spatial parameters in the complex octonion curved space
}


\author{Zi-Hua Weng
}


\institute{Zi-Hua Weng \at
              School of Physics and Mechanical \& Electrical Engineering, Xiamen University, Xiamen 361005, China    \\
              \email{xmuwzh@xmu.edu.cn}           
}

\date{Received: date / Accepted: date}

\maketitle

\begin{abstract}
The paper focuses on finding out several physical quantities to exert an influence on the spatial parameters of complex-octonion curved space, including the metric coefficient, connection coefficient, and curvature tensor. In the flat space described with the complex octonions, the radius vector is combined with the integrating function of field potential to become a composite radius vector. And the latter can be considered as the radius vector in a flat composite-space (a function space). Further it is able to deduce some formulae between the physical quantity and spatial parameter, in the complex-octonion curved composite-space. Under the condition of weak field approximation, these formulae infer a few results accordant with the General Theory of Relativity. The study reveals that it is capable of ascertaining which physical quantities are able to result in the warping of space, in terms of the curved composite-space described with the complex octonions. Moreover, the method may be expanded into some curved function spaces, seeking out more possible physical quantities to impact the bending degree of curved spaces.

\end{abstract}

\section{Introduction}

In the General Theory of Relativity (GR for short), A. Einstein claimed that the energy-momentum tensor will result in the warping of space. According to the Kerr and Schwarzschild vacuum solutions, it is able to explain a few physical phenomena in the gravitational field under the extreme conditions. The amazing achievements acquired in the GR inspirit scholars to endeavor to clarify more puzzles. For instance, are there other physical quantities that may lead to warping the space also? How many kinds of physical quantities may we find to be able to make a contribution to the curved space? These puzzles have been intriguing and bewildering the scholars since a long time. Until recently, the emergence of electromagnetic and gravitational theories \cite{weng1} , described with the complex octonions (or the standard octonions, with few coordinate values being complex numbers), replies to a part of these conundrums. In the paper, making use of the curved composite-space (in Section 3) described with the complex octonions, one can find that there are some physical quantities, relevant to the electromagnetic and gravitational fields, will exert an influence on the bending degree of curved space.

In 1843, W. R. Hamilton invented the quaternion. Subsequently the octonion was found by others. When a part of coordinate values of the quaternion and octonion are complex numbers, they are called as the complex-quaternion and complex-octonion respectively. J. C. Maxwell was the first to apply simultaneously the algebra of quaternions and the vector terminology to research the physical property of electromagnetic field. Nowadays some scholars introduce the quaternion and octonion to study the physical property of electromagnetic and gravitational fields, including their properties in the curved space.

In recent years, a few scholars adopt the quaternion and/or octonion to describe various physical properties in the curved space. C. G. Oliveira \emph{et al.} \cite{oliveira} considered the covariant differential properties of the split Cayley subalgebra of local real quaternion tetrads. M. Tanisli \emph{et al.} \cite{tanisli} introduced a new generalized complex octonionic field term consisting of electromagnetic and gravito-electromagnetic components. C. Castro \cite{castro} proposed a nonassociative octonionic ternary gauge field theories based on a ternary bracket. S. V. Ludkowski \emph{et al.} \cite{ludkowski} aimed at spectral theory of super-differential operators over the quaternion skew field and the octonion algebra. Pushpa \emph{et al.} \cite{pushpa} discussed the grand unified theories, in terms of quaternions and octonions by using the relation between quaternion basis elements with Pauli matrices and octonions. B. C. Chanyal \emph{et al.} \cite{chanyal1} made an attempt to reformulate the generalized field equation of dyons, in terms of octonion variables. Meanwhile the authors \cite{chanyal2} attempted to analyze the role of split octonions in various unified field theories associated with dyons and the dark matter. J. W. Moffat \cite{moffat} showed an eight-dimensional Riemannian geometry to be the basis of a nonsymmetric theory of gravitation. J. D. Edmonds \cite{edmonds} extended the quaternion formulation of relativistic quantum theory to include curvilinear coordinates and curved space-time. S. Marques-Bonham \cite{marques-bonham} developed the geometrical properties of a flat tangent space-time local to the manifold of the Einstein-Schr\"{o}dinger nonsymmetric theory, to which an internal octonionic space is attached. A. R. Dundarer \cite{dundarer} showed that a field satisfies a self-duality relation in eight-dimensional curved space, with the winding number of the mapping given by the octonionic transformation. C. G. Tsagas \cite{tsagas} considered the evolution of electromagnetic fields in curved spacetimes, calculating the exact wave equations for the associated electric and magnetic components.

Making use of some theoretical discussions, scholars anticipate that it is possible to figure out several problems remaining to be solved in the GR, including the equivalence principle, energy-momentum tensor, and other physical quantities to cause the warping of space. And even they expect the researches to be able to develop the GR, explaining more physical phenomena, such as the astrophysical jets \cite{weng12} , dark matter, and quantum phenomenon and so forth.

In 1915, A. Einstein published the GR. But so far, the scholars cannot yet deduce directly the exact solutions, which accord with any given initial and boundary conditions, from the Einstein's field equations and energy-momentum tensor. Undoubtedly, this embarrassing situation disturbs applying the GR to elucidate numerous physical phenomena. At present, there are a few exact solutions. However the effectual exact solutions, which possess the practical physical meanings and relate with the astronomical observations, are merely exiguous still, including the Schwarzschild vacuum solution in 1916 and the Kerr vacuum solution in 1962. As a result, scholars are suspicious of the GR from time to time, especially the energy-momentum tensor and equivalence principle. They engage in attempting to validate the equivalence principle from various aspects.

In the astronomical observations, some scholars survey the equivalence principle, from the Shapiro delay, dark matter, and microsatellite and so on. J.-J. Wei \emph{et al.} \cite{wei} found that the observed time delays between different energy bands from TeV blazars provide a way of testing the Einstein Equivalence Principle. N. Mohapi \emph{et al.} \cite{mohapi} considered a general tensor-scalar theory that allows testing the equivalence principle in the dark sector. F.-T. Han \emph{et al.} \cite{han} proposed a scheme for a space free-fall based test of the equivalence principle with two rotating extended bodies. R. D. Reasenberg \cite{reasenberg} described a new class of test masses for use in a Galilean test of the equivalence principle. H. Haghi \emph{et al.} \cite{haghi} revealed that the strong version of the equivalence principle is violated, while the gravitational dynamics of a system is influenced by the external gravitational field. R. Bousso \cite{bousso} found an infalling observer is unable to excite the vacuum near the black hole horizon, challenging the equivalence principle. J. Overduin \emph{et al.} \cite{overduin} applied observational uncertainties in the positions and motions of solar-system bodies, violating the equivalence principle. E. O. Kahya \emph{et al.} \cite{kahya} pointed out that GW150914 experienced a Shapiro delay, constraining any violations of equivalence principle. E. Hardy \emph{et al.} \cite{hardy} aimed at testing the Equivalence Principle via the precise measurement delivered by a differential electrostatic accelerometer on-board a drag-free microsatellite. C. J. A. P. Martins \emph{et al.} \cite{martins} discussed how the measurements constrain the simplest class of dynamical dark energy models, leading to violations of the Weak Equivalence Principle. W.-T. Ni \cite{ni} gave a fairly detailed account of the construction of the light cone and a core metric from the equivalence principle for photons. P. C. C. Freire \emph{et al.} \cite{freire} proposed a new strong equivalence principle violation test, based on the measurement of the variation of the orbital eccentricity.

In terms of the quantum particle, a part of scholars try to explore the equivalence principle, from the dual-species atom interferometers, dual-species matter-wave accelerometer, and International Space Station and so forth. B. Barrett \emph{et al.} demonstrated two methods \cite{barrett} for extracting the differential phase between dual-species atom interferometers for tests of the weak equivalence principle. A. Bonnin \emph{et al.} \cite{bonnin} presented the performance of a simultaneous dual-species matter-wave accelerometer for measuring their differential acceleration. J. F. Donoghue \emph{et al.} \cite{donoghue} discussed the long-distance corrections at one loop lead to quantum violations of some classical consequences of the equivalence principle. L. Zhou \emph{et al.} \cite{zhou} notified an improved test of the weak equivalence principle by using a simultaneous 85Rb-87Rb dual-species atom interferometer. M. G. Tarallo \emph{et al.} \cite{tarallo} reported on a test of the equivalence principle, measuring the acceleration in Earth's gravity field of two isotopes of strontium atoms. M. A. Hohensee \emph{et al.} \cite{hohensee} reported a joint test of local Lorentz invariance and the Einstein equivalence principle for electrons. H. Hernandez-Coronado \emph{et al.} \cite{hernandez-coronado} showed that the compatibility between the equivalence principle and quantum mechanics does not depend on the introduction of a mass superselection rule. C. Armendariz-Picon \emph{et al.} \cite{armendariz-picon} studied the equivalence principle and its violations by quantum effects in scalar-tensor theories that admit a conformal frame. S. V. Mousavi \emph{et al.} \cite{mousavi} researched the effect of quantum statistics on the arrival time distribution of quantum particles computed through the probability current density. J. Williams \emph{et al.} \cite{williams} described the quantum test of the equivalence principle and space time, a concept for an atom interferometry mission on the International Space Station. B. Altschul \emph{et al.} \cite{altschul} presented the scientific objectives in fundamental physics of the `space-time explorer and quantum equivalence space test' space mission. I. Licata \emph{et al.} \cite{licata} considered a direct coupling between the Ricci curvature scalar and the matter Lagrangian, and the acceleration is retrieved at low energies.

In the theoretical research, several scholars develop a few new theories, surveying and even violating the equivalence principle. N. E. J. Bjerrum-Bohr \emph{et al.} \cite{bjerrum-bohr} showed how modern methods can be applied to quantum gravity at low energy, challenging the classical framework behind the equivalence principle. A. Esmaili \emph{et al.} \cite{esmaili} analyzed the Ice Cube atmospheric neutrino data to constrain the violation of equivalence principle. S. T. Pereira \emph{et al.} \cite{pereira} observed that the compatibility of nonrelativistic quantum mechanics with Einstein's equivalence principle have been constrained to the existence of a superselection rule. Kh. P. Gnatenko \cite{gnatenko} concluded that the weak equivalence principle is violated in the case of a non-uniform gravitational field. S. Ghosh \cite{ghosh} showed that the Equivalence Principle is violated by Quantum Gravity effects. E. Di Casola \emph{et al.} \cite{casola} built a formal operative test able to probe the validity of the `gravitational weak equivalence principle' for any metric theory of gravity. V. M. Tkachuk \cite{tkachuk} noticed that the bodies of different mass fall in different ways in some spaces, and the equivalence principle is violated. F. M. Huber \emph{et al.} \cite{huber} proposed a test of the Weak Equivalence Principle for antiprotons in microgravity, expecting to test the Weak Equivalence Principle for antimatter to about one part per million. A. Saha \cite{saha} claimed that adding a mass-dependent contribution to the gravitational acceleration makes time-space noncommutativity a potential candidate for an apparent violation of the weak equivalence principle. C. S. Unnikrishnan \emph{et al.} \cite{unnikrishnan} wondered whether the interaction-induced inertia in the Higgs mechanism is the same as the charge of gravity or the gravitational mass, querying the weak equivalence principle. G. Lambiase \cite{lambiase} assumed that the gravitational coupling of neutrinos is flavor dependent, which implies a violation of the equivalence principle.

Making a comparison and analysis of preceding studies, a few primal problems associate with the GR are found as follows:

a) A controversial postulate. The E\"{o}tv\"{o}s experiment has never been validated under any ultra-strong field, especially the ultra-strong magnetic field, while a few theories predict that the gravitational mass would be variable. As a postulate derived from the E\"{o}tv\"{o}s experiment, the equivalence principle has been becoming a vulnerable point to be criticized. Some scholars criticize these relevant issues, from the unfinished experiments to the derived postulate and so forth. Now there are already several theories violating the equivalence principle \cite{damour} .
The suspicion keeps coming until now. Apparently this postulate should be inappropriate, and even unnecessary.

b) Single physical quantity. The viewpoint of `the field dominates the space' in the GR belongs to the groundbreaking insight, according with the Cartesian concept of `the space is the extension of substance'. The paper will succeed to this concept. But the GR is unable to express this extraordinary viewpoint adequately. Especially the theory does not take into account the feature that each fundamental field possesses various physical quantities, surmising that there is merely the `energy-momentum tensor' to be capable of impacting the warping of space. Obviously it is deficient.

c) Challenge of experiments. As an approximate theory of the GR, the Newtonian mechanics is incapable of explaining some physical phenomena, including the dark matter, astrophysical jets, and dark energy. Either the GR is unable to figure out these puzzles. These physical phenomena violate the theories. It means that the field equations in the GR are appealing for more validation experiments. So the scholars put the effort towards a few theoretical schemes, such as, `Beyond the Relativity', `Beyond the Standard Model', and `Superstring theory' and so forth.

Presenting a striking contrast to the above is that it is able to account for a few questions, derived from the GR, in the curved composite-space described with the complex octonions, trying to improve the academic thought of the GR to a certain extent.

a) Measurement of curvature. The scholars should not presume transcendentally that the physical space is curved or flat. The bending degree of curved space would depend on the measurement result of experiments. And even one may deem that the postulate derived from the E\"{o}tv\"{o}s experiment must be unnecessary. It can be expected that the fundamental assumption of the GR would be expressed more reasonably than ever, in case we consider over the requirement of `the measurement result determines the bending degree of curved space'.

b) Some physical quantities. According to the concept of R. Descartes, M. Faraday, and A. Einstein, the space is the extension of the fundamental field. In other words, the physical property of field dominates that of space. Each field possesses various physical quantities, while several physical quantities may have the possibility to exert an influence on the bending degree of curved space. In the paper, some physical quantities may impact the bending degree of curved space at different levels, including a few physical quantities similar to the `energy-momentum tensor'.

c) Scope of theory. In the flat space described with the complex octonions, the field theory consists of the Newtonian mechanics and Maxwell's field equations. And it is capable of explaining the astrophysical jets, dark matter, and precessional phenomenon and so forth. In the curved space described with the complex octonions, the gravitational and electromagnetic theories are expected to elucidate much more complicated physical phenomena than ever, in the curved spaces.

The paper will focus on discussing some possible physical quantities to result in the bending of space. Applying the curved space described with the complex octonions is capable of depicting relations between the physical quantity of fundamental fields with the spatial parameter of curved space. Especially a part of physical quantities may exert an influence on a few spatial parameters directly. On the basis of the composite radius vector (in Section 4) described with the complex octonions, it is found that some physical quantities associated with the composite radius vector will impact the spatial parameter, operator, and field equations and so on in the curved space. Under the condition of weak field approximation, it is able to deduce a few inferences accordant with the GR. Based on the complex-octonion angular momentum (in Section 5), the study reveals that some physical quantities related with the angular momentum can also affect the spatial parameter, operator, and field equations and so forth in the curved space.

\section{Curved space}

In the curved space described with the complex octonions (rather than the split octonions), by means of the definitions of manifold, affine frame, and metric space, it is able to deduce some spatial parameters under the orthogonal and unequal-length affine frame, including the metric coefficient, connection coefficient, covariant derivative, torsion tensor, and curvature tensor and so forth.

\subsection{Metric space}

The manifold is an indefinite geometric object. One can establish the coordinate systems in the manifold. By means of a continuous-differential function, it is able to convert a coordinate system into the other. The geometric structure of manifold is quite scrimpy. We have not found any other geometric properties in the manifold, except for the capability for transforming the coordinate systems. The manifold has neither metric tensor nor affine connection. That is, the tensor in the manifold does not possess the metric property, while there is no relation among the tensors of different points.

It is allowed to define firstly the connection among different tensors in contiguous points, even if there is no metric property in the manifold. And this manifold with the connection is called the affine connection space. Besides the connection coefficient, an affine connection space may possess the curvature tensor and torsion tensor. In the paper, we merely discuss a simple case the torsion tensor is equal to zero. That is, the connection coefficient, $\Gamma_{\alpha\beta}^\gamma$ , is symmetric with respect to the subscripts, $\alpha$ and $\beta$ . After obtaining the connection coefficient, it is able to define the parallel translation among the tensors.

In a few manifolds, there may be the property of `parallel translation' among the tensors of different points. When a tensor transfers from one point to another to meet the requirement of parallel translation, it means that the differential of this tensor is equal to zero. According to this definition, we can achieve the relations between the tensor component and connection coefficient, in terms of parallel translation. Obviously the parallel translation is a geometric property irrespective of the choice of coordinate systems. In compliance with the definition of tensor, it is found that the connection coefficient is not a tensor.

After achieving the affine connection among the tensors, it is able to define the metric property in the affine connection space. Now we obtain one metric space to possess not only parallel translation but also metric property, including the Riemannian and pseudo-Riemannian spaces. In the curved space described with the complex octonions, it is capable of defining the tangent-frame component, by means of the partial derivative of complex-octonion tensor with respect to the coordinate value. According to the definition of octonion norm, the metric coefficient can be derived from the tangent-frame component, and is octonion-Hermitian \cite{weng49} .

In the metric space, there are few relations between the connection coefficient and metric coefficient, enabling these two kinds of coefficients to be consistent with each other. The connection coefficient can be derived from the metric coefficient, while the curvature tensor is inferred from the connection coefficient. In the curved space, after experiencing the parallel translation, the result of one tensor correlates with not only the starting and ending positions but also the movement paths. In virtue of the existence of curvature tensor, the parallel translations of one tensor may be different from each other, when the tensor experiences different movement paths.

In the complex-octonion curved space $\mathbb{O}$ , the partial derivative of radius vector with respect to the coordinate value can be utilized for the component of affine frame in the tangent space. Apparently, choosing an appropriate affine frame will be beneficial to lower down the difficulty of mathematics may be encountered in the physics theories. a) Orthogonality. The affine frame should be orthogonal, enabling the metric coefficient and connection coefficient both to be scalar, reducing the mathematical difficulty related with the nonassociativity of octonions in the following context. b) Normalization. The orthogonal affine frame may be unequal-length. Further, it is necessary to convert the orthogonal and unequal-length affine frame into the orthogonal and equal-length affine frame, extending the physical laws from the flat space into the orthogonal and equal-length affine frame in the tangent space. c) Bending degree. The bending degree of curved space will impact the physical laws in the orthogonal and unequal-length affine frame in the tangent space. Making use of contrasting the measurement result with theoretical prediction in the tangent space, it is capable of estimating the bending degree of the curved space.

\subsection{Curvature tensor}

In the flat space described with the complex octonions, it is proper to choose the rectangular coordinate system, in which the complex-octonion radius vector is,
\begin{equation}
\mathbb{H} ( h^\alpha ) = i h^0 \textbf{\emph{i}}_0 + h^r \textbf{\emph{i}}_r + i h^4 \textbf{\emph{i}}_4 + h^{4+r} \textbf{\emph{i}}_{4+r}  ~ ,
\end{equation}
where $h^\alpha$ is the coordinate value. $h^j$ is real, while $h^{j+4}$ is the imaginary number, to meet the requirement of the dimensional homogeneity. $\textbf{\emph{i}}_\alpha$ is the basis vector. $\textbf{\emph{i}}_0 = 1$. $\textbf{\emph{i}}_0^2 = 1$. $\textbf{\emph{i}}_\xi^2 = -1$. $j = 0, 1, 2, 3$. $r, q = 1, 2, 3$. $ \alpha , \beta , \gamma , \lambda , \nu = 0, 1, 2, 3, 4, 5, 6, 7$. $\xi = 1, 2, 3, 4, 5, 6, 7$.

In the complex-octonion curved space, one tangent space may be considered as a locally flat space. Therefore the physical laws and relevant research methods can be extended from the complex-octonion flat space into the tangent space. The rectangular coordinate system in the flat space will be transformed into the orthogonal and equal-length affine frame in the tangent space, further converting into the orthogonal and unequal-length affine frame.

In the tangent space, one orthogonal and unequal-length tangent frame is utilized for the coordinate system, in which the complex-octonion radius vector is expanded in terms of the tangent-frame component $\textbf{\emph{e}}_\alpha$ ,
\begin{equation}
\mathbb{H} ( c^\alpha ) = i c^0 \textbf{\emph{e}}_0 + c^r \textbf{\emph{e}}_r + i c^4 \textbf{\emph{e}}_4 + c^{4+r} \textbf{\emph{e}}_{4+r}  ~ ,
\end{equation}
where $c^\alpha$ is the coordinate value. $c^j$ is real, while $c^{j+4}$ is the imaginary number, to meet the requirement of the dimensional homogeneity. $\textbf{\emph{e}}_\alpha$ is unequal-length. Choosing an appropriate coordinate system to satisfy the conditions that, $\textbf{\emph{e}}_0$ is the scalar part (corresponding to $\textbf{\emph{i}}_0$ ), $\textbf{\emph{e}}_r$ is the component of vector part (corresponding to $\textbf{\emph{i}}_r$ ), $\textbf{\emph{e}}_4$ is the `scalar' part (corresponding to $\textbf{\emph{i}}_4$ ), and $\textbf{\emph{e}}_{4+r}$ is the component of `vector' part (corresponding to $\textbf{\emph{i}}_{4+r}$ ). $\textbf{\emph{e}}_0^2 > 0$ , and $\textbf{\emph{e}}_\xi^2 < 0$ .

Making use of the tangent-frame component and the norm of complex-octonion radius vector, it is able to define the metric of complex-octonion curved space $\mathbb{O}$ ,
\begin{equation}
d S^2 = g_{\overline{\alpha} \beta} d\overline{u^\alpha} du^\beta ~ ,
\end{equation}
where the metric coefficient, $g_{\overline{\alpha} \beta} = \textbf{\emph{e}}_\alpha^* \circ \textbf{\emph{e}}_\beta$ , is octonion-Hermitian. $\textbf{\emph{e}}_\alpha$ is the component of tangent frame, with $\textbf{\emph{e}}_\alpha = \partial \mathbb{H} / \partial u^\alpha $ . $\ast$ denotes the octonion conjugate. $u^0 = i c^0$ , $u^r = c^r$ , $u^4 = i c^4$ , $u^{4+r} = c^{4+r}$ . $( u^\alpha )^* = \overline{u^\alpha}$ , and it indicates that the correlated tangent-frame component, $\textbf{\emph{e}}_\alpha$ , is octonion conjugate. $g_{\overline{\alpha} \beta}$ is scalar, due to the orthogonal tangent frame.

From the metric coefficient, one can deduce the connection coefficient (Appendix A), in the complex-octonion curved space,
\begin{eqnarray}
\Gamma_{\overline{\lambda} , \beta \gamma } = (1/2) ( \partial g_{\overline{\gamma} \lambda} / \partial u^\beta + \partial g_{\overline{\lambda} \beta} / \partial u^\gamma - \partial g_{\overline{\gamma} \beta} /\partial u^\lambda ) ~ ,
\\
\Gamma_{\overline{\lambda} , \overline{\beta} \gamma } = (1/2) ( \partial g_{\overline{\gamma} \lambda} / \partial \overline{u^\beta} + \partial g_{\overline{\lambda} \beta} / \partial \overline{u^\gamma} - \partial g_{\overline{\gamma} \beta} /\partial \overline{u^\lambda} ) ~ ,
\end{eqnarray}
where $\Gamma_{\overline{\lambda} , \beta \gamma }$ and $\Gamma_{\overline{\lambda} , \overline{\beta} \gamma }$ both are connection coefficients, and are all scalar. $\Gamma_{\overline{\lambda} , \beta \gamma } = g_{\overline{\lambda} \alpha} \Gamma_{\beta \gamma}^\alpha$ . $ \Gamma_{\beta \gamma}^\alpha = g^{\alpha \overline{\lambda}} \Gamma_{\overline{\lambda} , \beta \gamma } $ . $\Gamma_{\overline{\lambda} , \overline{\beta} \gamma } = g_{\overline{\lambda} \alpha} \Gamma_{\overline{\beta} \gamma}^\alpha$ . $ \Gamma_{\overline{\beta} \gamma}^\alpha =  g^{\alpha \overline{\lambda}} \Gamma_{\overline{\lambda} , \overline{\beta} \gamma } $ . $\Gamma_{\beta \gamma}^\alpha = \Gamma_{\gamma \beta}^\alpha $ . $\Gamma_{\overline{\beta} \gamma}^\alpha = \Gamma_{\gamma \overline{\beta}}^\alpha $ . $ g^{\alpha \overline{\lambda}}  g_{\overline{\lambda} \beta} = \delta^\alpha_\beta $ . $[ ( \Gamma_{\overline{\beta} \gamma}^\alpha )^* ]^T = \Gamma_{\gamma \overline{\beta}}^\alpha $ . The superscript T denotes the transpose of matrix.

When a complex-octonion quantity, $\mathbb{Y} = Y^\alpha \textbf{\emph{e}}_\alpha $ , is transferred from a point $M_1$ to the next point $M_2$ , to meet the requirement of parallel translation, it states that the differential of quantity $\mathbb{Y}$ equals to zero. And the condition of parallel translation, $d \mathbb{Y} = 0$ , will yield,
\begin{equation}
d Y^\beta = - \Gamma_{\alpha \gamma}^\beta Y^\alpha d u^\gamma  ~,
\end{equation}
with
\begin{equation}
\partial^2 \mathbb{H} / \partial u^\beta \partial u^\gamma = \Gamma_{\beta \gamma}^\alpha \textbf{\emph{e}}_\alpha  ~ .
\end{equation}

In the complex-octonion curved space, for the first-rank contravariant tensor $Y^\beta$ of a point $M_2$ , the component of covariant derivative with respect to the coordinate $u^\gamma$ is written as,
\begin{eqnarray}
\nabla_\gamma Y^\beta = \partial ( \delta_\alpha^\beta Y^\alpha ) / \partial u^\gamma + \Gamma_{\alpha \gamma}^\beta Y^\alpha   ~ ,
\\
\nabla_{\overline{\gamma}} Y^\beta = \partial ( \delta_\alpha^\beta Y^\alpha ) / \partial \overline{u^\gamma} + \Gamma_{\alpha \overline{\gamma}}^\beta Y^\alpha   ~ ,
\end{eqnarray}
where $Y^\beta$ is scalar.

Next, from the covariant derivative of tensor, it is able to infer the curvature tensor (Appendix B), in the complex-octonion curved space,
\begin{equation}
\nabla_{\overline{\alpha}} ( \nabla_\beta Y^\gamma ) - \nabla_\beta ( \nabla_{\overline{\alpha}} Y^\gamma ) = R_{\beta \overline{\alpha} \nu}^{~~~~~~\gamma} Y^\nu
+ T_{\beta \overline{\alpha}}^\lambda ( \nabla_\lambda Y^\gamma )  ~   ,
\end{equation}
with
\begin{equation}
R_{\beta \overline{\alpha} \nu}^{~~~~~~\gamma} = \partial \Gamma_{\nu \beta}^\gamma / \partial \overline{u^\alpha} - \partial \Gamma_{\nu \overline{\alpha}}^\gamma  / \partial u^\beta
+ \Gamma_{\lambda \overline{\alpha}}^\gamma \Gamma_{\nu \beta}^\lambda - \Gamma_{\lambda \beta}^\gamma \Gamma_{\nu \overline{\alpha}}^\lambda   ~~ ,
\end{equation}
where $R_{\beta \overline{\alpha} \nu}^{~~~~~~\gamma}$ and $T_{\beta \overline{\alpha}}^\lambda$ both are scalar. $R_{\beta \overline{\alpha} \nu}^{~~~~~~\gamma}$ is the curvature tensor, while $T_{\beta \overline{\alpha}}^\lambda$ is the torsion tensor. $T_{\beta \overline{\alpha}}^\lambda = \Gamma_{\overline{\alpha} \beta }^\lambda - \Gamma_{\beta \overline{\alpha}}^\lambda $ . In the paper, we merely discuss the case, $T_{\beta \overline{\alpha}}^\lambda = 0$, that is, $\Gamma_{\overline{\alpha} \beta }^\lambda = \Gamma_{\beta \overline{\alpha}}^\lambda $ .

When some equations of electromagnetic and gravitational fields are extended from the complex-octonion flat space into the complex-octonion curved space, the covariant derivative and curvature tensor will make a contribution to these equations to a certain extent. It is the research we have discussed in Ref.[51], which is very noticeably different from the paper. We place great emphasis on the influence of several physical quantities of electromagnetic and gravitational fields on the spatial parameters of complex-octonion curved space in the paper.

However, in the complex-octonion curved space, it is incapable of establishing directly the relationships between the physical quantities and spatial parameters. By contrast, in the complex-octonion curved composite-space, it is able to deduce the formulae between the physical quantities and spatial parameters, achieving a few inferences in accordance with that of the GR. Apparently the research methods of the complex-octonion curved space in the above can be extended into that of the curved composite-space described with the complex octonions.

\begin{table}[h]
\caption{Several equations associated with the gravitational and electromagnetic fields in the complex-octonion curved space $\mathbb{O}$ . The spatial parameters of curved space $\mathbb{O}$
make a contribution towards the operator $\lozenge$ , impacting the field equations (See Ref.[51]). $k_{rx}$ , $\mu$ , and $\mu_g$ are coefficients. }
\label{tab:table3}
\centering
\begin{tabular}{ll}
\hline\hline
physics~quantity             &   definition                                                                                 \\
\hline
field~potential              &  $\mathbb{A} = i \lozenge^\star \circ \mathbb{X}  $                                          \\
field~strength               &  $\mathbb{F} = \lozenge \circ \mathbb{A}  $                                                  \\
field~source                 &  $\mu \mathbb{S} = - ( i \mathbb{F} / v_0 + \lozenge )^* \circ \mathbb{F} $                  \\
linear~momentum              &  $\mathbb{P} = \mu \mathbb{S} / \mu_g $                                                      \\
composite~radius~vector      &  $\mathbb{U} = \mathbb{H} + k_{rx} \mathbb{X}  $                                             \\
angular~momentum             &  $\mathbb{L} = \mathbb{U}^\star \circ \mathbb{P} $                                           \\
\hline\hline
\end{tabular}
\end{table}

\section{Composite space}

In the complex-octonion flat space, the complex octonion can be applied to describe the electromagnetic and gravitational fields (see Ref.[1]), deducing the field potential, field strength, field source, angular momentum, torque, and force and so forth, inferring the current continuity equation, mass continuity equation, and precession equation, and even depicting the astrophysical jets (see Ref.[14]) and dark matter and so on. Either of the radius vector $\mathbb{H}$ and the integrating function $\mathbb{X}$ of field potential will play a non-ignorable role in these theoretical descriptions.

In case neglecting the contribution of the integrating function $\mathbb{X}$ of field potential in the composite radius vector, $\mathbb{U} = \mathbb{H} + k_{rx} \mathbb{X}$ , the energy expression would lack for the `electromagnetic potential energy' and `gravitational potential energy' and so forth. As a result, the composite radius vector, $\mathbb{U} = \mathbb{H} + k_{rx} \mathbb{X}$ , should be considered as a whole to take into account in the following context. Further the composite radius vector can be regarded as one radius vector in the function space, which is called as the composite space temporarily. Herein $k_{rx}$ is a coefficient, to meet the requirement of the dimensional homogeneity.

In the flat composite space described with the complex octonions, it is able to choose the rectangular coordinate system, in which the composite radius vector is written as,
\begin{equation}
\mathbb{U} ( H^\alpha ) = i H^0 \textbf{\emph{i}}_0 + H^r \textbf{\emph{i}}_r + i H^4 \textbf{\emph{i}}_4 + H^{4+r} \textbf{\emph{i}}_{4+r}  ~ ,
\end{equation}
where $H^\alpha$ is the coordinate value, with $H^\alpha = h^\alpha + k_{rx} x^\alpha$ . The integrating function of field potential is, $\mathbb{X} ( x^\alpha ) = i x^0 \textbf{\emph{i}}_0 + x^r \textbf{\emph{i}}_r + i x^4 \textbf{\emph{i}}_4 + x^{4+r} \textbf{\emph{i}}_{4+r}$ . $H^j$ and $x^j$ are all real. $H^{j+4}$ and $x^{j+4}$ both are imaginary numbers.

In the curved composite-space $\mathbb{O}_U$ described with the complex octonions, the orthogonal and unequal-length tangent frame is utilized for the coordinate system in the tangent space. In the coordinate system of the tangent space, the composite radius vector can be expanded in terms of the tangent-frame component $\textbf{\emph{E}}_\alpha$ ,
\begin{equation}
\mathbb{U} ( C^\alpha ) = i C^0 \textbf{\emph{E}}_0 + C^r \textbf{\emph{E}}_r + i C^4 \textbf{\emph{E}}_4 + C^{4+r} \textbf{\emph{E}}_{4+r}  ~ ,
\end{equation}
where $C^\alpha$ and $X^\alpha$ both are coordinate values, with $C^\alpha = c^\alpha + k_{rx} X^\alpha$ . The integrating function of field potential is, $\mathbb{X} ( X^\alpha ) = i X^0 \textbf{\emph{E}}_0 + X^r \textbf{\emph{E}}_r + i X^4 \textbf{\emph{E}}_4 + X^{4+r} \textbf{\emph{E}}_{4+r}$ . $\textbf{\emph{E}}_\alpha$ is unequal-length. Choosing an appropriate coordinate system to meet the demand that, $\textbf{\emph{E}}_0$ is the scalar part, $\textbf{\emph{E}}_r$ is the component of vector part, while $\textbf{\emph{E}}_4$ is the `scalar' part, and $\textbf{\emph{E}}_{4+r}$ is the component of `vector' part. $C_j$ and $X_j$ are all real. $C_{j+4}$ and $X_{j+4}$ both are imaginary numbers.  $\textbf{\emph{E}}_0^2 > 0$ , and $\textbf{\emph{E}}_\xi^2 < 0$ .

From the point of view of function space, the composite space is one kind of function space. Consequently it is allowed to apply the research methods of function space to study the physical properties relevant to the composite space. The function space may comprise multiple physical properties, and its coordinate value is a function. In other words, the coordinate value $C^\alpha$ of composite space can be regarded as one function, while the coordinate value $C^\alpha$ consists of the spatial parameter $c^\alpha$ and physical quantity $X^\alpha$ . When the composite space is distorted, it is found that there are some other formulae between the spatial parameter and physical quantity.

In the composite space, there is, $\mathbb{H} \gg k_{rx} \mathbb{X}$ , in general, that is, $\mathbb{H} + k_{rx} \mathbb{X} \approx \mathbb{H}$ . So it is not easy to perceive the possible influence of the integrating function $\mathbb{X}$ of field potential on the composite radius vector in most situations. However, as the derivative of the integrating function of field potential, the field potential, field strength, energy, and refractive index and so forth will play an important role in the field theories. It reveals that the integrating function of field potential is indispensable.

\begin{table}[h]
\caption{ Some major composite arguments may be decoupled, in the composite space $\mathbb{O}_U$ described with the complex octonions. Under some special circumstances, each of composite arguments could be separated into the spatial parameter, physical quantity, and coupling term. }
\label{tab:table3}
\centering
\begin{tabular}{llll}
\hline\hline
composite~argument                                           &   spatial~parameter                                          &   physical~quantity                         &   coupling~term        \\
\hline
$g_{\overline{\alpha} \beta (H,X)}$                          &  $g_{\overline{\alpha} \beta (H)}$                           &   $g_{\overline{\alpha} \beta (X)}$
                                                             &  $c.t.(  g_{\overline{\alpha} \beta} , \mathbb{H} , \mathbb{X} )$                                                                   \\
$\Gamma_{\alpha \beta (H,X)}^\gamma$                         &  $\Gamma_{\alpha \beta (H)}^\gamma$                          &   $\Gamma_{\alpha \beta (X)}^\gamma$
                                                             &  $c.t.( \Gamma_{\alpha \beta }^\gamma , \mathbb{H} , \mathbb{X} )$                                                                  \\
$\Gamma_{\overline{\alpha} \beta (H,X)}^\gamma$              &  $\Gamma_{\overline{\alpha} \beta (H)}^\gamma$               &   $\Gamma_{\overline{\alpha} \beta (X)}^\gamma$
                                                             &  $c.t.(  \Gamma_{\overline{\alpha} \beta }^\gamma , \mathbb{H} , \mathbb{X} )$                                                      \\
$R_{\alpha \overline{\beta} \nu \overline{\lambda} (H,X)}$   &  $R_{\alpha \overline{\beta} \nu \overline{\lambda} (H)}$    &   $R_{\alpha \overline{\beta} \nu \overline{\lambda} (X)}$
                                                             &  $c.t.( R_{\alpha \overline{\beta} \nu \overline{\lambda} } , \mathbb{H} , \mathbb{X} ) $                                           \\
\hline\hline
\end{tabular}
\end{table}

\section{Composite radius vector}

In the curved composite-space described with the complex octonions, the partial derivative of composite radius vector can be utilized for the component of tangent frame. Obviously, this is a new kind of curved space, on the basis of the composite radius vector. And the underlying space is the complex-octonion composite space, while the tangent space is the complex-octonion composite space as well. By means of the physical properties of composite radius vector, it is capable of deducing some arguments, including the metric coefficient, connection coefficient, covariant derivative, and curvature tensor and so forth. According to the decomposing of curvature tensor, it is found some relations between the spatial parameter and physical quantity.

\subsection{Curved composite space}

In the curved composite-space described with the complex octonions, the partial derivative $\textbf{\emph{E}}_\alpha$ of composite radius vector with respect to the coordinate value $U^\alpha$ is used for the component of tangent frame in the tangent space. These orthogonal and unequal-length components of tangent frame comprise one coordinate system, ensuring certain arguments to be scalar, including the metric coefficient, connection coefficient, and curvature tensor. From the component of tangent frame, it is able to define the metric of curved composite-space $\mathbb{O}_U$ , described with the complex octonions,
\begin{equation}
d S^2_{(H,X)} = g_{\overline{\alpha} \beta (H,X)} d \overline{U^\alpha} d U^\beta  ~ ,
\end{equation}
where the metric tensor is, $g_{\overline{\alpha} \beta (H,X)} = \textbf{\emph{E}}_\alpha^\ast \circ \textbf{\emph{E}}_\beta $ , with $g_{\overline{\alpha} \beta (H,X)}$ being the scalar. $\textbf{\emph{E}}_\alpha$ is the component of tangent frame, with $\textbf{\emph{E}}_\alpha = \partial \mathbb{U} / \partial U^\alpha $ . $U^0 = i C^0$ , $U^r = C^r$ , $U^4 = i C^4$ , and $U^{4+r} = C^{4+r}$ . And $( U^\alpha )^\ast = \overline{U^\alpha}$ , while it indicates that the correlated component, $\textbf{\emph{E}}_\alpha$ , is octonion-conjugate.

In the curved composite-space $\mathbb{O}_U$ , substituting the coordinate value $U^\alpha$ , tangent-frame component $\textbf{\emph{E}}_\alpha$ , and metric coefficient $g_{\overline{\alpha} \beta (H,X)}$ for the coordinate value $u^\alpha$ , tangent-frame component $\textbf{\emph{e}}_\alpha$ , and metric coefficient $g_{\overline{\alpha} \beta}$ in the complex-octonion curved space $\mathbb{O}$ respectively, it is capable of inferring the connection coefficients of the curved composite-space $\mathbb{O}_U$ . And there are,
\begin{eqnarray}
\Gamma_{\overline{\lambda} , \beta \gamma  (H,X)} = (1/2) ( \partial g_{\overline{\gamma} \lambda (H,X)} / \partial U^\beta + \partial g_{\overline{\lambda} \beta (H,X)} / \partial U^\gamma - \partial g_{\overline{\gamma} \beta (H,X)} /\partial U^\lambda ) ~ ,
\\
\Gamma_{\overline{\lambda} , \overline{\beta} \gamma  (H,X)} = (1/2) ( \partial g_{\overline{\gamma} \lambda (H,X)} / \partial \overline{U^\beta} + \partial g_{\overline{\lambda} \beta (H,X)} / \partial \overline{U^\gamma} - \partial g_{\overline{\gamma} \beta (H,X)} /\partial \overline{U^\lambda} ) ~ ,
\end{eqnarray}
where $\Gamma_{\overline{\lambda} , \beta \gamma  (H,X)}$ and $\Gamma_{\overline{\lambda} , \overline{\beta} \gamma  (H,X)}$ both are connection coefficients, and are all scalar. $\Gamma_{\overline{\lambda} , \beta \gamma  (H,X)} = g_{\overline{\lambda} \alpha (H,X)} \Gamma_{\beta \gamma (H,X)}^\alpha$ . $ \Gamma_{\beta \gamma (H,X)}^\alpha = g^{\alpha \overline{\lambda}}_{ (H,X)} \Gamma_{\overline{\lambda} , \beta \gamma  (H,X)} $ . $\Gamma_{\beta \gamma (H,X)}^\alpha = \Gamma_{\gamma \beta (H,X)}^\alpha $ . $ \Gamma_{\overline{\beta} \gamma (H,X)}^\alpha =  g^{\alpha \overline{\lambda}}_{ (H,X)} \Gamma_{\overline{\lambda} , \overline{\beta} \gamma  (H,X)} $ . $\Gamma_{\overline{\lambda} , \overline{\beta} \gamma  (H,X)} = g_{\overline{\lambda} \alpha (H,X)} \Gamma_{\overline{\beta} \gamma (H,X)}^\alpha$ . $\Gamma_{\overline{\beta} \gamma (H,X)}^\alpha = \Gamma_{\gamma \overline{\beta} (H,X)}^\alpha $ . $ g^{\alpha \overline{\lambda}}_{ (H,X)}  g_{\overline{\lambda} \beta (H,X)} = \delta^\alpha_\beta $ . $[ ( \Gamma_{\overline{\beta} \gamma (H,X)}^\alpha )^* ]^T = \Gamma_{\gamma \overline{\beta} (H,X)}^\alpha $ .

In the curved composite-space described with the complex octonions, when the physical quantity, $\mathbb{Y} = Y^\beta \textbf{\emph{E}}_\beta $ , is transferred from a point $M_1$ to the next point $M_2$ , to meet the requirement of parallel translation, it means that the differential of quantity $\mathbb{Y}$ equals to zero. This condition of parallel translation, $d \mathbb{Y} = 0$, will infer,
\begin{equation}
d Y^\beta = - \Gamma_{\alpha \gamma (H,X)}^\beta Y^\alpha d U^\gamma  ~,
\end{equation}
with
\begin{equation}
\partial^2 \mathbb{U} / \partial U^\beta \partial U^\gamma = \Gamma_{\beta \gamma (H,X)}^\alpha \textbf{\emph{E}}_\alpha  ~ .
\end{equation}

For the first-rank contravariant tensor $Y^\beta$ of a point $M_2$ , in the curved composite-space described with the complex octonions, the component of covariant derivative with respect to the coordinate $U^\gamma$ is written as,
\begin{eqnarray}
\nabla_\gamma Y^\beta = \partial ( \delta_\alpha^\beta Y^\alpha ) / \partial U^\gamma + \Gamma_{\alpha \gamma (H,X)}^\beta Y^\alpha   ~ ,
\\
\nabla_{\overline{\gamma}} Y^\beta = \partial ( \delta_\alpha^\beta Y^\alpha ) / \partial \overline{U^\gamma} + \Gamma_{\alpha \overline{\gamma} (H,X)}^\beta Y^\alpha   ~ ,
\end{eqnarray}
where $Y^\beta$ is scalar.

From the property of covariant derivative of tensor, the definition of curvature tensor is,
\begin{equation}
\nabla_{\overline{\alpha}} ( \nabla_\beta Y^\gamma ) - \nabla_\beta ( \nabla_{\overline{\alpha}} Y^\gamma ) = R_{\beta \overline{\alpha} \nu ~~(H,X)}^{~~~~~~\gamma} Y^\nu
+ T_{\beta \overline{\alpha} (H,X)}^\lambda ( \nabla_\lambda Y^\gamma )  ~   ,
\end{equation}
with
\begin{eqnarray}
R_{\beta \overline{\alpha} \nu ~~(H,X)}^{~~~~~~\gamma} = && \partial \Gamma_{\nu \beta (H,X)}^\gamma / \partial \overline{U^\alpha} - \partial \Gamma_{\nu \overline{\alpha} (H,X)}^\gamma  / \partial U^\beta
\nonumber
\\
&& + \Gamma_{\lambda \overline{\alpha} (H,X)}^\gamma \Gamma_{\nu \beta (H,X)}^\lambda - \Gamma_{\lambda \beta (H,X)}^\gamma \Gamma_{\nu \overline{\alpha} (H,X)}^\lambda   ~~ ,
\end{eqnarray}
where $R_{\beta \overline{\alpha} \nu ~~(H,X)}^{~~~~~~\gamma}$ and $T_{\beta \overline{\alpha} (H,X)}^\lambda$ both are scalar. $R_{\beta \overline{\alpha} \nu ~~(H,X)}^{~~~~~~\gamma}$ is the curvature tensor, while $T_{\beta \overline{\alpha} (H,X)}^\lambda$ is the torsion tensor. $T_{\beta \overline{\alpha} (H,X)}^\lambda = \Gamma_{\overline{\alpha} \beta (H,X)}^\lambda - \Gamma_{\beta \overline{\alpha} (H,X)}^\lambda $ . In the paper, we only discuss the case, $T_{\beta \overline{\alpha} (H,X)}^\lambda = 0$, that is, $\Gamma_{\overline{\alpha} \beta (H,X)}^\lambda = \Gamma_{\beta \overline{\alpha} (H,X)}^\lambda $ .

By means of the analysis of metric coefficient, connection coefficient, and curvature tensor and so forth, it is found that the spatial parameter and physical quantity both make a contribution towards these arguments, in the curved composite-space described with the complex octonions. In case the coupling of spatial parameters and physical quantities could be decoupled in these arguments, it may be able to seek out some direct interrelations between the spatial parameter and physical quantity.

\subsection{Coupling term}

From the preceding equations, it is found that the radius vector $\mathbb{H}$ and the integrating function $\mathbb{X}$ of field potential are linearly superposed, in the curved composite-space. This is a comparatively simple case. Nevertheless, as the application of the derivative of composite radius vector, the interrelations between the spatial parameters and physical quantities in the following context will be enough complicated, including certain interconnections among the metric coefficient, connection coefficient, and curvature tensor. The coupling term (c.t. for short) between the spatial parameter and physical quantity is always highly intricate, and it is very tough to decouple in general. Only in some special cases, it may be decoupled approximately, and then we stand a chance of looking into a few explicit relations of spatial parameters with physical quantities.

From the metric equation, Eq.(14), it is found that the contributions, coming from the spatial parameter and physical quantity, to the metric coefficient are closely interrelated. Both of partial derivatives, $ \partial \mathbb{H} / \partial U^\alpha $ and $ \partial \mathbb{X} / \partial U^\alpha $ , will exert an influence on the metric coefficient. Meanwhile the term, $ \partial \mathbb{X} / \partial U^\alpha $ , possesses the dimension of field potential (Appendix C). In terms of the metric tensor, $g_{\overline{\alpha} \beta (H,X)}$ , when the contribution of $\mathbb{X}$ could be neglected, $g_{\overline{\alpha} \beta (H,X)}$ will be reduced into $g_{\overline{\alpha} \beta (H)}$ , which contains only the contribution of $ \partial \mathbb{H} / \partial U^\alpha $ . Similarly, if the contribution of $\mathbb{H}$ could be neglected, $g_{\overline{\alpha} \beta (H,X)}$ will be reduced into $g_{\overline{\alpha} \beta (X)}$ , which contains only the contribution of $ \partial \mathbb{X} / \partial U^\alpha $ . Consequently, in a few comparatively simple cases, the metric tensor can be separated into,
\begin{equation}
g_{\overline{\alpha} \beta (H,X)} = g_{\overline{\alpha} \beta (H)} + g_{\overline{\alpha} \beta (X)} + c.t.( g_{\overline{\alpha} \beta} , \mathbb{H} , \mathbb{X} )   ~   .
\end{equation}

In the connection coefficient, Eq.(15), it is found that the contributions, coming from the spatial parameter and physical quantity, to the connection coefficient are mingled together. Both of partial derivatives, $ \partial g_{\overline{\alpha} \beta (H)} / \partial U^\gamma $ and $ \partial g_{\overline{\alpha} \beta (X)} / \partial U^\gamma  $ , will have an influence on the connection coefficient. And the term, $(k_{rx}^{~~-1}) \partial g_{\overline{\alpha} \beta (X)} / \partial U^\gamma  $, possesses the dimension of field strength. For the connection coefficient, $\Gamma_{\alpha \beta (H,X)}^\gamma$ , when the contribution of $g_{\overline{\alpha} \beta (X)}$ could be neglected, $\Gamma_{\alpha \beta (H,X)}^\gamma$ will be reduced into $\Gamma_{\alpha \beta (H)}^\gamma$ , which contains only the contribution of $ \partial g_{\overline{\alpha} \beta (H)} / \partial U^\gamma  $ . In a similar way, in case the contribution of $g_{\overline{\alpha} \beta (H)}$ could be neglected, $\Gamma_{\alpha \beta (H,X)}^\gamma$ will be reduced into $\Gamma_{\alpha \beta (X)}^\gamma$ , which contains only the contribution of $ \partial g_{\overline{\alpha} \beta (X)} / \partial U^\gamma  $ . As a result, in several comparatively simple cases, the connection coefficient can be divided into,
\begin{equation}
\Gamma_{\alpha \beta (H,X)}^\gamma = \Gamma_{\alpha \beta (H)}^\gamma + \Gamma_{\alpha \beta (X)}^\gamma + c.t.( \Gamma_{\alpha \beta}^\gamma , \mathbb{H} , \mathbb{X} )   ~   .
\end{equation}

The decomposition process of connection coefficient, Eq.(16), is similar to that of Eq.(15), in some comparatively simple cases.

The analysis of the curvature tensor, Eq.(22), shows that the contributions, coming from the spatial parameter and physical quantity, to the curvature tensor are mixed complicatedly. Four partial derivatives, $ \partial \Gamma_{\alpha \beta (H)}^\gamma / \partial U^\nu $ , $ \partial \Gamma_{\alpha \beta (X)}^\gamma / \partial U^\nu $ , $ \partial \Gamma_{\overline{\alpha} \beta (H)}^\gamma / \partial U^\nu $, and $ \partial \Gamma_{\overline{\alpha} \beta (X)}^\gamma / \partial U^\nu $ , will directly impact the curvature tensor. And both of terms,
 $ (k_{rx}^{~~-1}) \partial \Gamma_{\alpha \beta (X)}^\gamma / \partial U^\nu $ and $ (k_{rx}^{~~-1}) \partial \Gamma_{\overline{\alpha} \beta (X)}^\gamma / \partial U^\nu $ ,
will possess the dimension of field source. In terms of the curvature tensor, $R_{\beta \overline{\alpha} \nu ~~(H,X)}^{~~~~~~\gamma}$ , when the contribution of $\Gamma_{\alpha \beta (X)}^\gamma$ and $\Gamma_{\overline{\alpha} \beta (X)}^\gamma$ both could be neglected, $R_{\beta \overline{\alpha} \nu ~~(H,X)}^{~~~~~~\gamma}$ will be reduced into $R_{\beta \overline{\alpha} \nu ~~(H)}^{~~~~~~\gamma}$ , which contains the contributions of $ \partial \Gamma_{\alpha \beta (H)}^\gamma / \partial U^\nu $ and $ \partial \Gamma_{\overline{\alpha} \beta (H)}^\gamma / \partial U^\nu $ . Similarly, when the contribution of $\Gamma_{\alpha \beta (H)}^\gamma$ and $\Gamma_{\overline{\alpha} \beta (H)}^\gamma$ both could be neglected, $R_{\beta \overline{\alpha} \nu ~~(H,X)}^{~~~~~~\gamma}$ will be reduced into $R_{\beta \overline{\alpha} \nu ~~(X)}^{~~~~~~\gamma}$ , which contains the contribution of $ \partial \Gamma_{\alpha \beta (X)}^\gamma / \partial U^\nu $ and $ \partial \Gamma_{\overline{\alpha} \beta (X)}^\gamma / \partial U^\nu $ . So, in some comparatively simple cases, the curvature tensor can be separated into,
\begin{equation}
R_{\beta \overline{\alpha} \nu ~~(H,X)}^{~~~~~~\gamma} = R_{\beta \overline{\alpha} \nu ~~(H)}^{~~~~~~\gamma} + R_{\beta \overline{\alpha} \nu ~~(X)}^{~~~~~~\gamma} + c.t.( R_{\beta \overline{\alpha} \nu }^{~~~~~~\gamma} , \mathbb{H} , \mathbb{X} )   ~   .
\end{equation}

It is well known that the physics is an experimental science substantially. Only by means of the precise experimental measurement, can we conclude whether the composite space is distorted. That is, whether the curvature tensor will be equal to zero depends on the measurement results. It implies that it is unnecessary to presume whether the composite space is distorted before the experiment measurement.

In certain circumstances, by means of the physical experiments, it is able to measure the bending degree of composite space to verify if it was flat. In other words, there may be the flat composite space under some cases, that is, $R_{\beta \overline{\alpha} \nu ~~(H,X)}^{~~~~~~\gamma} = 0$ . It means that the curvature tensor $R_{\beta \overline{\alpha} \nu ~~(H,X)}^{~~~~~~\gamma}$ is still possible to be zero, even if neither of ingredients, $R_{\beta \overline{\alpha} \nu ~~(H)}^{~~~~~~\gamma}$ and $R_{\beta \overline{\alpha} \nu ~~(X)}^{~~~~~~\gamma}$ , is equal to zero. Undeniably there must be a few relations between these two ingredients.

When $R_{\beta \overline{\alpha} \nu ~~(H,X)}^{~~~~~~\gamma} = 0$, one simple formula among the ingredients of curvature tensor in the flat composite space can be written as,
\begin{equation}
0 = R_{\beta \overline{\alpha} \nu ~~(H)}^{~~~~~~\gamma} + R_{\beta \overline{\alpha} \nu ~~(X)}^{~~~~~~\gamma} + c.t.( R_{\beta \overline{\alpha} \nu }^{~~~~~~\gamma} , \mathbb{H} , \mathbb{X} )   ~   ,
\end{equation}
further there may be, $c.t.( R_{\beta \overline{\alpha} \nu }^{~~~~~~\gamma} , \mathbb{H} , \mathbb{X} ) = 0$. In case the coupling term is equal to zero, the above can be reduced into one simpler formula,
\begin{equation}
R_{\beta \overline{\alpha} \nu ~~(H)}^{~~~~~~\gamma} = - R_{\beta \overline{\alpha} \nu ~~(X)}^{~~~~~~\gamma}   ~   ,
\end{equation}
where the left side is the contribution of the spatial parameter, while the right side the physical quantity. The term $(k_{rx}^{~~-1}) R_{\beta \overline{\alpha} \nu ~~(X)}^{~~~~~~\gamma}$ possesses the dimension of field source, while its ingredients are much more complicated than that of field source.

The above consists with the Einsteinian academic thought, that is, the existence of field dominates the bending of space. Further the above intuitively opens out this academic thought from another point of view. Certainly the GR and the paper both succeed to Cartesian academic thought of `the space is the extension of substance'.

In some other circumstances, making use of the physical experiments, it is capable of measuring the bending degree of composite space to examine if it was curved. In other words, there may be the curved composite-space under some other cases, that is, $R_{\beta \overline{\alpha} \nu ~~(H,X)}^{~~~~~~\gamma} \neq 0$ . It means that either of ingredients, $R_{\beta \overline{\alpha} \nu ~~(H)}^{~~~~~~\gamma}$ and $R_{\beta \overline{\alpha} \nu ~~(X)}^{~~~~~~\gamma}$ , may not be equal to zero, while there must be a few complicated coupling relations between these two ingredients. In this case, the solution of the component of curvature tensor depends on the difficulty of curvature tensor, Eq.(22), in a particular question. Certainly it will be tough to resolve.

In the curved composite-space described with the complex octonions, when the contributions of some physical quantities are comparatively huge, it will exert an evident influence on the spatial parameters of curved space $\mathbb{O}$ , including the metric coefficient, connection coefficient, and curvature tensor. From the above, when the contribution of the `physical quantity', $ \partial \mathbb{X} / \partial U^\alpha $ , is large, it will result in altering of the metric coefficient, impacting the connection coefficient and curvature tensor. Similarly, when the contribution of the `physical quantities', $ \partial g_{\overline{\alpha} \beta (X)} / \partial U^\gamma $ or $ \partial \Gamma_{\alpha \beta (X)}^\gamma / \partial U^\nu $ , is big, it will lead to changing of the curvature tensor.

After achieving the spatial parameters, one can find that these arguments will have an important influence on certain operators, impacting some equations of electromagnetic and gravitational fields. On the basis of composite radius vector $\mathbb{U}$ , some `physical quantities', $g_{\gamma \overline{\lambda} (X)}$ , $\Gamma_{\beta \nu (X)}^\lambda$ , $\Gamma_{\overline{\beta} \nu (X)}^\lambda$ , and $R_{\beta \overline{\alpha} \nu \overline{\lambda} (X)}$ , may make a contribution towards the spatial parameters, $g_{\gamma \overline{\lambda} (H)}$ , $\Gamma_{\beta \nu (H)}^\lambda$ , $\Gamma_{\overline{\beta} \nu (H)}^\lambda$ , and $R_{\beta \overline{\alpha} \nu \overline{\lambda} (H)}$ . Due to, $\mathbb{H} + k_{rx} \mathbb{X} \approx \mathbb{H}$ , in general, these spatial parameters will approximate to that in Section 2. Subsequently, these spatial parameters will exert an influence on certain operators in the curved space $\mathbb{O}$ , such as, divergence, gradient, and curl. Finally, these operators is capable of impacting directly some equations of electromagnetic and gravitational fields in the curved space $\mathbb{O}$ . All of a sudden, it is found that the study of curved space $\mathbb{O}$ (and the GR) is so closely interrelated with the electromagnetic and gravitational fields described with the complex octonions, by means of the composite space.

In the complex-octonion flat space, not all the partial derivatives of any one physical quantity possess a practical physical meaning. In contrast, the appropriate combination of some partial derivatives is possible to be seized of certain practical physical meanings, including the divergence, gradient, and curl. Further, in the complex-octonion curved space $\mathbb{O}$ , the partial derivative of any one physical quantity may not become a new physical quantity with the practical physical meaning either. Nevertheless, the proper combination of some covariant derivatives of physical quantities may form several new physical quantities with the practical physical meanings, including the divergence, gradient, and curl of physical quantities.

In the composite space, the integrating function, $\mathbb{X}$ , of field potential is a component of the composite radius vector, and is a physical quantity. Also the connection coefficient and curvature tensor correlative with this physical quantity become two new physical quantities, including $\Gamma_{\alpha \beta (X)}^\gamma$ and $R_{\beta \overline{\alpha} \nu ~~(X)}^{~~~~~~\gamma}$ . By means of the composite space $\mathbb{O}_U$ , the two new physical quantities dominate directly two spatial parameters, $\Gamma_{\alpha \beta (H)}^\gamma$ and $R_{\beta \overline{\alpha} \nu ~~(H)}^{~~~~~~\gamma}$ respectively. Therefore these new physical quantities are seized of the practical physical meanings. It should be noted that these new physical quantities are different from the physical quantities in Ref.[51].

The research method in the curved space $\mathbb{O}_U$ , based on the composite radius vector, can be extended into other curved spaces based on other arguments (such as, angular momentum). In these curved spaces, some new physical quantities will be possible to dominate the spatial parameters of curved spaces, impacting some equations of electromagnetic and gravitational fields from different standpoints. It means that there may be several curved spaces on the basis of different arguments. Further these different curved spaces may be capable of exerting an influence on the some equations of electromagnetic and gravitational fields simultaneously.

\begin{table}[h]
\caption{Certain primary composite arguments may be decoupled, in the angular-momentum space $\mathbb{O}_L$ described with the complex octonions. Under some particular situations, each of composite arguments would be divided into the spatial parameter, physical quantity, and coupling term. }
\label{tab:table3}
\centering
\begin{tabular}{llll}
\hline\hline
composite~argument                                             &   spatial~parameter                                            &   physical~quantity                         &   coupling~term     \\
\hline
$g_{\overline{\alpha} \beta (P,H,X)}$                          &  $g_{\overline{\alpha} \beta (P,H)}$                           &   $g_{\overline{\alpha} \beta (P,X)}$
                                                               &  $c.t.(  g_{\overline{\alpha} \beta} , \mathbb{P} , \mathbb{H} , \mathbb{X} )$                                                     \\
$\Gamma_{\alpha \beta (P,H,X)}^\gamma$                         &  $\Gamma_{\alpha \beta (P,H)}^\gamma$                          &   $\Gamma_{\alpha \beta (P,X)}^\gamma$
                                                               &  $c.t.( \Gamma_{\alpha \beta }^\gamma , \mathbb{P} , \mathbb{H} , \mathbb{X} )$                                                    \\
$\Gamma_{\overline{\alpha} \beta (P,H,X)}^\gamma$              &  $\Gamma_{\overline{\alpha} \beta (P,H)}^\gamma$               &   $\Gamma_{\overline{\alpha} \beta (P,X)}^\gamma$
                                                               &  $c.t.(  \Gamma_{\overline{\alpha} \beta }^\gamma , \mathbb{P} , \mathbb{H} , \mathbb{X} )$                                        \\
$R_{\alpha \overline{\beta} \nu \overline{\lambda} (P,H,X)}$   &  $R_{\alpha \overline{\beta} \nu \overline{\lambda} (P,H)}$    &   $R_{\alpha \overline{\beta} \nu \overline{\lambda} (P,X)}$
                                                               &  $c.t.( R_{\alpha \overline{\beta} \nu \overline{\lambda} } , \mathbb{P} , \mathbb{H} , \mathbb{X} ) $                             \\
\hline\hline
\end{tabular}
\end{table}

\section{Angular momentum}

In the curved composite-space described with the complex octonions, the partial derivative of complex-octonion angular momentum can be utilized for the component of tangent frame in the tangent space. Apparently, this is a new kind of curved space or curved function space, on the basis of the complex-octonion angular momentum. And it is called as the curved angular-momentum space temporarily. The underlying space and tangent space both are the angular-momentum spaces described with the complex octonions. By means of the physical properties of complex-octonion angular momentum, it is capable of deducing some kinds of arguments, including the metric coefficient, connection coefficient, covariant derivative, and curvature tensor and so on. From the decomposing of curvature tensor, one can shed light on some relations between the spatial parameter and physical quantity. From a different perspective, it is revealed that there are some other physical quantities will exert an influence on the spatial parameters, in the curved angular-momentum space.

\subsection{Curved angular-momentum space}

In the curved angular-momentum space $\mathbb{O}_L$ described with the complex octonions, the partial derivative $\textbf{\emph{E}}_{\alpha (P,H,X)}$ of complex-octonion angular momentum with respect to the coordinate value is utilized for the component of tangent frame in the tangent space. These orthogonal and unequal-length components of tangent frame comprise a coordinate system, ensuring a few arguments to be scalar, including the metric coefficient, connection coefficient, and curvature tensor. In the coordinate system of tangent space, the complex-octonion angular momentum $\mathbb{L}$ can be expanded in terms of the tangent-frame component, $\textbf{\emph{E}}_{\alpha (P,H,X)}$ , and written as, $\mathbb{L} = L^\alpha \textbf{\emph{E}}_{\alpha (P,H,X)}$ . From the component of tangent frame, it is able to define the metric of curved angular-momentum space $\mathbb{O}_L$ , described with the complex octonions,
\begin{equation}
d S^2_{(P,H,X)} = g_{\overline{\alpha} \beta (P,H,X)} d \overline{L^\alpha} d L^\beta  ~ ,
\end{equation}
where the metric tensor is, $g_{\overline{\alpha} \beta (P,H,X)} = \textbf{\emph{E}}_{\alpha (P,H,X)}^\ast \circ \textbf{\emph{E}}_{\beta (P,H,X)} $ . $L^\alpha$ is chosen as the coordinate value. The tangent-frame component is, $\textbf{\emph{E}}_{\beta (P,H,X)} = \partial \mathbb{L} / \partial L^\beta $ . The complex-octonion angular momentum is, $ \mathbb{L} = ( \mathbb{H} + k_{rx} \mathbb{X} )^\star \circ \mathbb{P} $ , while $\mathbb{P}$ is the complex-octonion linear momentum. $\star$ denotes the complex conjugate. $\textbf{\emph{E}}_{\alpha (P,H,X)}$ is unequal-length. Choosing an appropriate coordinate system to meet the requirement that, $\textbf{\emph{E}}_{0 (P,H,X)}$ is the scalar part, $\textbf{\emph{E}}_{r (P,H,X)}$ is the component of vector part, meanwhile $\textbf{\emph{E}}_{4 (P,H,X)}$ is the `scalar' part, and $\textbf{\emph{E}}_{4+r (P,H,X)}$ is the component of `vector' part. $g_{\overline{\alpha} \beta (P,H,X)}$ and $L^\alpha$ both are scalar.  $\textbf{\emph{E}}_{0 (P,H,X)}^2 > 0$ , and $\textbf{\emph{E}}_{\xi (P,H,X)}^2 < 0$ .

In the curved angular-momentum space described with the complex octonions, substituting the coordinate value $L^\alpha$ , tangent-frame component $\textbf{\emph{E}}_{\alpha (P,H,X)}$ , and metric coefficient $g_{\overline{\alpha} \beta (P,H,X)}$ for the coordinate value $U^\alpha$ , tangent-frame component $\textbf{\emph{E}}_{\alpha}$ , and metric coefficient $g_{\overline{\alpha} \beta (H,X)}$ in the curved composite-space $\mathbb{O}_U$ respectively, it is capable of inferring the connection coefficients of the curved angular-momentum space $\mathbb{O}_L$ . And there exist,
\begin{eqnarray}
\Gamma_{\overline{\lambda} , \beta \gamma  (P,H,X)} = && (1/2) ( \partial g_{\overline{\gamma} \lambda (P,H,X)} / \partial L^\beta + \partial g_{\overline{\lambda} \beta (P,H,X)} / \partial L^\gamma
\nonumber
\\
&&
~~~~~~~~~~~ - \partial g_{\overline{\gamma} \beta (P,H,X)} /\partial L^\lambda ) ~ ,
\\
\Gamma_{\overline{\lambda} , \overline{\beta} \gamma  (P,H,X)} = && (1/2) ( \partial g_{\overline{\gamma} \lambda (P,H,X)} / \partial \overline{L^\beta} + \partial g_{\overline{\lambda} \beta (P,H,X)} / \partial \overline{L^\gamma}
\nonumber
\\
&&
~~~~~~~~~~~ - \partial g_{\overline{\gamma} \beta (P,H,X)} /\partial \overline{L^\lambda} ) ~ ,
\end{eqnarray}
where $\Gamma_{\overline{\lambda} , \beta \gamma  (P,H,X)}$ and $\Gamma_{\overline{\lambda} , \overline{\beta} \gamma  (P,H,X)}$ both are connection coefficients, and are scalar. $\Gamma_{\overline{\lambda} , \beta \gamma  (P,H,X)} = g_{\overline{\lambda} \alpha (P,H,X)} \Gamma_{\beta \gamma (P,H,X)}^\alpha$ . $\Gamma_{\overline{\lambda} , \overline{\beta} \gamma  (P,H,X)} = g_{\overline{\lambda} \alpha (P,H,X)} \Gamma_{\overline{\beta} \gamma (P,H,X)}^\alpha$ . $ \Gamma_{\beta \gamma (P,H,X)}^\alpha = g^{\alpha \overline{\lambda}}_{ (P,H,X)} \Gamma_{\overline{\lambda} , \beta \gamma  (P,H,X)} $ . $ \Gamma_{\overline{\beta} \gamma (P,H,X)}^\alpha =  g^{\alpha \overline{\lambda}}_{ (P,H,X)} \Gamma_{\overline{\lambda} , \overline{\beta} \gamma  (P,H,X)} $ . $[ ( \Gamma_{\overline{\beta} \gamma (P,H,X)}^\alpha )^* ]^T = \Gamma_{\gamma \overline{\beta} (P,H,X)}^\alpha $ . $ g^{\alpha \overline{\lambda}}_{ (P,H,X)}  g_{\overline{\lambda} \beta (P,H,X)} = \delta^\alpha_\beta $ . $\Gamma_{\beta \gamma (P,H,X)}^\alpha = \Gamma_{\gamma \beta (P,H,X)}^\alpha $ . $\Gamma_{\overline{\beta} \gamma (P,H,X)}^\alpha = \Gamma_{\gamma \overline{\beta} (P,H,X)}^\alpha $ .

In the curved angular-momentum space described with the complex octonions, when a physical quantity, $\mathbb{Y} = Y^\beta \textbf{\emph{E}}_{\beta (P,H,X)}$ , is transferred from a point $M_1$ to the next point $M_2$ , to meet the demand of parallel translation, it means that the differential of quantity $\mathbb{Y}$ equals to zero. This condition of parallel translation, $d \mathbb{Y} = 0$, will deduce,
\begin{equation}
d Y^\beta = - \Gamma_{\alpha \gamma (P,H,X)}^\beta Y^\alpha d L^\gamma  ~,
\end{equation}
with
\begin{equation}
\partial^2 \mathbb{L} / \partial L^\beta \partial L^\gamma = \Gamma_{\beta \gamma (P,H,X)}^\alpha \textbf{\emph{E}}_{\alpha (P,H,X)}  ~ .
\end{equation}

For the first-rank contravariant tensor $Y^\beta$ of a point $M_2$ , in the curved angular-momentum space described with the complex octonions, the component of covariant derivative with respect to the coordinate $L^\gamma$ is written as,
\begin{eqnarray}
\nabla_\gamma Y^\beta = \partial ( \delta_\alpha^\beta Y^\alpha ) / \partial L^\gamma + \Gamma_{\alpha \gamma (P,H,X)}^\beta Y^\alpha   ~ ,
\\
\nabla_{\overline{\gamma}} Y^\beta = \partial ( \delta_\alpha^\beta Y^\alpha ) / \partial \overline{L^\gamma} + \Gamma_{\alpha \overline{\gamma} (P,H,X)}^\beta Y^\alpha   ~ ,
\end{eqnarray}
where $Y^\beta$ is scalar.

According to the property of covariant derivative of tensor, the definition of curvature tensor is written as,
\begin{equation}
\nabla_{\overline{\alpha}} ( \nabla_\beta Y^\gamma ) - \nabla_\beta ( \nabla_{\overline{\alpha}} Y^\gamma ) = R_{\beta \overline{\alpha} \nu ~~(P,H,X)}^{~~~~~~\gamma} Y^\nu
+ T_{\beta \overline{\alpha} (P,H,X)}^\lambda ( \nabla_\lambda Y^\gamma )  ~   ,
\end{equation}
with
\begin{eqnarray}
R_{\beta \overline{\alpha} \nu ~~(P,H,X)}^{~~~~~~\gamma} = && \partial \Gamma_{\nu \beta (P,H,X)}^\gamma / \partial \overline{L^\alpha} - \partial \Gamma_{\nu \overline{\alpha} (P,H,X)}^\gamma  / \partial L^\beta
\nonumber
\\
&& + \Gamma_{\lambda \overline{\alpha} (P,H,X)}^\gamma \Gamma_{\nu \beta (P,H,X)}^\lambda - \Gamma_{\lambda \beta (P,H,X)}^\gamma \Gamma_{\nu \overline{\alpha} (P,H,X)}^\lambda   ~~ ,
\end{eqnarray}
where $R_{\beta \overline{\alpha} \nu ~~(P,H,X)}^{~~~~~~\gamma}$ and $T_{\beta \overline{\alpha} (P,H,X)}^\lambda$ both are scalar. $R_{\beta \overline{\alpha} \nu ~~(P,H,X)}^{~~~~~~\gamma}$ is the curvature tensor, while $T_{\beta \overline{\alpha} (P,H,X)}^\lambda$ is the torsion tensor. $T_{\beta \overline{\alpha} (P,H,X)}^\lambda = \Gamma_{\overline{\alpha} \beta (P,H,X)}^\lambda - \Gamma_{\beta \overline{\alpha} (P,H,X)}^\lambda $ . In the paper, we only discuss the case, $T_{\beta \overline{\alpha} (P,H,X)}^\lambda = 0$, that is, $\Gamma_{\overline{\alpha} \beta (P,H,X)}^\lambda = \Gamma_{\beta \overline{\alpha} (P,H,X)}^\lambda $ .

Making use of the analysis of metric coefficient, connection coefficient, and curvature tensor, one can find that the spatial parameter and physical quantity both exert an influence on these arguments, in the curved angular-momentum space described with the complex octonions. In case the coupling of spatial parameters and physical quantities could be decoupled in these arguments, it is possible to glimpse at several connections of spatial parameter with physical quantity. Certainly it must be more complicated than that in the curved composite-space.

\subsection{Coupling term}

From the above, it is found that the `spatial parameter' $\mathbb{H}^\star \circ \mathbb{P}$ and `physical quantity' $\mathbb{X}^\star \circ \mathbb{P}$ can be linearly superposed, in the curved angular-momentum space. These two arguments are similar to that in Section 4, but they are not unmixed spatial parameter or physical quantity. As the application of the derivative of these two arguments, the relations between the spatial parameter and physical quantity in the following context are quite complicated, including some relations among the metric coefficient, connection coefficient, and curvature tensor. The coupling terms among these arguments are often complicated, and it is tough to decouple in general. Only in some special situations, it may be decoupled approximately. The challenge is finding ways to decouple relevant arguments in a particular question.

From the metric equation, Eq.(28), it is found that the contributions, coming from two terms, $\mathbb{H}^\star \circ \mathbb{P}$ and $\mathbb{X}^\star \circ \mathbb{P}$ , to the metric coefficient are closely interconnected. Both of partial derivatives, $\partial ( \mathbb{H}^\star \circ \mathbb{P} ) / \partial L^\alpha $ and $\partial ( \mathbb{X}^\star \circ \mathbb{P} ) / \partial L^\alpha $ , will exert an influence on the metric coefficient. Meanwhile the term, $\partial ( \mathbb{X}^\star \circ \mathbb{P} ) / \partial L^\alpha $ , possesses the dimension of field potential. In terms of the metric tensor, $g_{\overline{\alpha} \beta (P,H,X)}$ , when the contribution of $\mathbb{X}^\star \circ \mathbb{P}$ could be neglected, $g_{\overline{\alpha} \beta (P,H,X)}$ will be reduced into $g_{\overline{\alpha} \beta (P,H)}$ , which contains only the contribution of $\partial ( \mathbb{H}^\star \circ \mathbb{P} ) / \partial L^\alpha $ . Similarly, if the contribution of $\mathbb{H}^\star \circ \mathbb{P}$ could be neglected, $g_{\overline{\alpha} \beta (P,H,X)}$ will be reduced into $g_{\overline{\alpha} \beta (P,X)}$ , which contains only the contribution of $\partial ( \mathbb{X}^\star \circ \mathbb{P} ) / \partial L^\alpha $ . As a result, in a few comparatively simple cases, the metric tensor can be separated into,
\begin{equation}
g_{\overline{\alpha} \beta (P,H,X)} = g_{\overline{\alpha} \beta (P,H)} + g_{\overline{\alpha} \beta (P,X)} + c.t.( g_{\overline{\alpha} \beta} , \mathbb{P} , \mathbb{H} , \mathbb{X} )   ~   .
\end{equation}

In the connection coefficient, Eq.(29), it is found that the contributions, coming from the two components, $g_{\overline{\alpha} \beta (P,H)}$ and $g_{\overline{\alpha} \beta (P,X)}$ , to the connection coefficient are mingled together. Both of partial derivatives, $\partial g_{\overline{\alpha} \beta (P,H)} / \partial L^\gamma $ and $\partial g_{\overline{\alpha} \beta (P,X)} / \partial L^\gamma $, have an influence on the connection coefficient. The term,$(k_{rx}^{~~-1}) \partial g_{\overline{\alpha} \beta (P,X)} / \partial L^\gamma $ , possesses the dimension of field strength. In terms of the connection coefficient, $\Gamma_{\alpha \beta (P,H,X)}^\gamma$ , when the contribution of $g_{\overline{\alpha} \beta (P,X)}$ could be neglected, $\Gamma_{\alpha \beta (P,H,X)}^\gamma$ will be reduced into $\Gamma_{\alpha \beta (P,H)}^\gamma$ , which contains only the contribution of $\partial g_{\overline{\alpha} \beta (P,H)} / \partial L^\gamma $ . In a similar way, in case the contribution of $g_{\overline{\alpha} \beta (P,H)}$ could be neglected, $\Gamma_{\alpha \beta (P,H,X)}^\gamma$ will be reduced into $\Gamma_{\alpha \beta (P,X)}^\gamma$ , which contains only the contribution of $\partial g_{\overline{\alpha} \beta (P,X)} / \partial L^\gamma $ . Therefore, in several comparatively simple cases, the connection coefficient can be divided into,
\begin{equation}
\Gamma_{\alpha \beta (P,H,X)}^\gamma = \Gamma_{\alpha \beta (P,H)}^\gamma + \Gamma_{\alpha \beta (P,X)}^\gamma + c.t.( \Gamma_{\alpha \beta}^\gamma , \mathbb{P} , \mathbb{H} , \mathbb{X} )   ~   .
\end{equation}

The decomposition process of connection coefficient, Eq.(30), is similar to that of Eq.(29), in some comparatively simple cases.

The analysis of the curvature tensor, Eq.(36), shows that the contributions, coming from the `spatial parameter' and `physical quantity', to the curvature tensor are mixed complicatedly. So four partial derivatives, $ \partial \Gamma_{\alpha \beta (P,H)}^\gamma / \partial L^\nu $ , $ \partial \Gamma_{\alpha \beta (P,X)}^\gamma / \partial L^\nu $ , $ \partial \Gamma_{\overline{\alpha} \beta (P,H)}^\gamma / \partial L^\nu $, and $ \partial \Gamma_{\overline{\alpha} \beta (P,X)}^\gamma / \partial L^\nu $ , will impact the curvature tensor directly. And both of terms, $(k_{rx}^{~~-1}) \partial \Gamma_{\alpha \beta (P,X)}^\gamma / \partial L^\nu $ and $(k_{rx}^{~~-1}) \partial \Gamma_{\overline{\alpha} \beta (P,X)}^\gamma / \partial L^\nu $ , will possess the dimension of field source. In terms of the curvature tensor, $R_{\beta \overline{\alpha} \nu ~~(P,H,X)}^{~~~~~~\gamma}$ , when the contribution of $\Gamma_{\alpha \beta (P,X)}^\gamma$ and $\Gamma_{\overline{\alpha} \beta (P,X)}^\gamma$ both could be neglected, $R_{\beta \overline{\alpha} \nu ~~(P,H,X)}^{~~~~~~\gamma}$ will be reduced into $R_{\beta \overline{\alpha} \nu ~~(P,H)}^{~~~~~~\gamma}$ , which contains the contributions of $ \partial \Gamma_{\alpha \beta (P,H)}^\gamma / \partial L^\nu $ and $ \partial \Gamma_{\overline{\alpha} \beta (P,H)}^\gamma / \partial L^\nu $ . Similarly, when the contribution of $\Gamma_{\alpha \beta (P,H)}^\gamma$ and $\Gamma_{\overline{\alpha} \beta (P,H)}^\gamma$ both could be neglected, $R_{\beta \overline{\alpha} \nu ~~(P,H,X)}^{~~~~~~\gamma}$ will be reduced into $R_{\beta \overline{\alpha} \nu ~~(P,X)}^{~~~~~~\gamma}$ , which contains the contribution of $ \partial \Gamma_{\alpha \beta (P,X)}^\gamma / \partial L^\nu $ and $ \partial \Gamma_{\overline{\alpha} \beta (P,X)}^\gamma / \partial L^\nu $ . Therefore, in a few comparatively simple cases, the curvature tensor can be separated into,
\begin{equation}
R_{\beta \overline{\alpha} \nu ~~(P,H,X)}^{~~~~~~\gamma} = R_{\beta \overline{\alpha} \nu ~~(P,H)}^{~~~~~~\gamma} + R_{\beta \overline{\alpha} \nu ~~(P,X)}^{~~~~~~\gamma} + c.t.( R_{\beta \overline{\alpha} \nu }^{~~~~~~\gamma} , \mathbb{P} , \mathbb{H} , \mathbb{X} )   ~   .
\end{equation}

In some circumstances, making use of the physical experiments, one can measure the bending degree of angular-momentum space to check if it was flat. In other words, there may be the flat angular-momentum space under some cases, that is, $R_{\beta \overline{\alpha} \nu ~~(P,H,X)}^{~~~~~~\gamma} = 0$. It means that the curvature tensor $R_{\beta \overline{\alpha} \nu ~~(P,H,X)}^{~~~~~~\gamma}$ is still possible to be zero, even if either of two ingredients, $R_{\beta \overline{\alpha} \nu ~~(P,H)}^{~~~~~~\gamma}$ and $R_{\beta \overline{\alpha} \nu ~~(P,X)}^{~~~~~~\gamma}$ , is not equal to zero. Undoubtedly there may be several relations between these two ingredients.

When $R_{\beta \overline{\alpha} \nu ~~(P,H,X)}^{~~~~~~\gamma} = 0$, a simple formula among the ingredients of curvature tensor in the flat angular-momentum space is written as,
\begin{equation}
0 = R_{\beta \overline{\alpha} \nu ~~(P,H)}^{~~~~~~\gamma} + R_{\beta \overline{\alpha} \nu ~~(P,X)}^{~~~~~~\gamma} + c.t.( R_{\beta \overline{\alpha} \nu }^{~~~~~~\gamma} , \mathbb{P} , \mathbb{H} , \mathbb{X} )   ~   ,
\end{equation}
further, in case the coupling term is equal to zero, there is, $c.t.( R_{\beta \overline{\alpha} \nu }^{~~~~~~\gamma} , \mathbb{P} , \mathbb{H} , \mathbb{X} ) = 0$, and then the above can be reduced into one simpler formula,
\begin{equation}
R_{\beta \overline{\alpha} \nu ~~(P,H)}^{~~~~~~\gamma} = - R_{\beta \overline{\alpha} \nu ~~(P,X)}^{~~~~~~\gamma}   ~   ,
\end{equation}
where the left side is the contribution of the `spatial parameter', $\mathbb{H}^\star \circ \mathbb{P}$ , while the right side the `physical quantity', $\mathbb{X}^\star \circ \mathbb{P}$ . The term $(k_{rx}^{~~-1}) R_{\beta \overline{\alpha} \nu ~~(P,X)}^{~~~~~~\gamma}$ is possessed of the dimension of field source, while its ingredients are much more complicated than that of field source.

Apparently, the above is still accordant with the academic thought of `the existence of field dominates the bending of space'. Further the above develops this academic thought. It points out that some other physical quantities are also possible to exert an influence on the bending degree of curved space, besides the physical quantities in Section 4.

In some other circumstances, by means of the physical experiments, it is able to measure the bending degree of angular-momentum space to inspect if it was curved. In other words, there may be the curved angular-momentum space under some other cases, that is, $R_{\beta \overline{\alpha} \nu ~~(P,H,X)}^{~~~~~~\gamma} \neq 0$. It states that two terms, $R_{\beta \overline{\alpha} \nu ~~(P,H)}^{~~~~~~\gamma}$ and $R_{\beta \overline{\alpha} \nu ~~(P,X)}^{~~~~~~\gamma}$, both may not be equal to zero, while there must be several intricate coupling relations between these two ingredients. In this situation, the solution of the component of curvature tensor lies on the difficulty of curvature tensor, Eq.(36), in a particular question. Of course it will be untoward to resolve.

In the curved angular-momentum space described with the complex octonions, when the contributions of some `physical quantities', $g_{\overline{\alpha} \beta (P,X)}$ , $\Gamma_{\alpha \beta (P,X)}^\gamma$ , $\Gamma_{\overline{\alpha} \beta (P,X)}^\gamma$ , and $R_{\beta \overline{\alpha} \nu ~~(P,X)}^{~~~~~~\gamma}$ , are comparatively huge, they will exert an evident influence on the `spatial parameters', $g_{\overline{\alpha} \beta (P,H)}$ , $\Gamma_{\alpha \beta (P,H)}^\gamma$ , $\Gamma_{\overline{\alpha} \beta (P,H)}^\gamma$ , and $R_{\beta \overline{\alpha} \nu ~~(P,H)}^{~~~~~~\gamma}$ , of curved space. Subsequently, these `spatial parameters' will make a contribution to some operators in the curved space, including the divergence, gradient, and curl. Finally the operators are able to impact directly some equations of electromagnetic and gravitational fields in the curved space.

Making a comparison and analysis of the preceding studies, it is found that there are a few discrepancies between two curved function spaces, $\mathbb{O}_L$ and $\mathbb{O}_U$ . a) Tangent frame. In the curved space $\mathbb{O}_U$ based on the composite radius vector, the tangent frame is associated with the composite radius vector $\mathbb{U}$ . In the curved space $\mathbb{O}_L$ based on the complex-octonion angular momentum, the tangent frame is relevant to the complex-octonion angular momentum $\mathbb{L}$ . b) Physical quantity. In the curved space based on the composite radius vector, the physical quantity is correlative with the integrating function, $\mathbb{X}$ , of field potential, impacting the spatial parameter of curved composite-space. In the curved space based on the complex-octonion angular momentum, the `physical quantity' is related to the component, $( \mathbb{X}^\star \circ \mathbb{P} )$ , affecting the `spatial parameter' of curved angular-momentum space. c) Spatial parameter. In the curved space $\mathbb{O}_U$ , the spatial parameters are associated with the radius vector $\mathbb{H}$, including the arguments, $g_{\overline{\alpha} \beta (H)}$ , $\Gamma_{\alpha \beta (H)}^\gamma$ , $\Gamma_{\overline{\alpha} \beta (H)}^\gamma$ , and $R_{\beta \overline{\alpha} \nu ~~(H)}^{~~~~~~\gamma}$ . In the curved space $\mathbb{O}_L$ , the `spatial parameter' are relevant to the component, $( \mathbb{H}^\star \circ \mathbb{P} )$ , including the arguments, $g_{\overline{\alpha} \beta (P,H)}$ , $\Gamma_{\alpha \beta (P,H)}^\gamma$ , $\Gamma_{\overline{\alpha} \beta (P,H)}^\gamma$ , and $R_{\beta \overline{\alpha} \nu ~~(P,H)}^{~~~~~~\gamma}$ .

To be brief, making use of the curved function space, it is able to catch sight of several relations between the `spatial parameters' and `physical quantities', in the curved space $\mathbb{O}_L$ described with the complex octonions. And these `spatial parameters' will make a contribution towards some operators of complex-octonion curved space $\mathbb{O}$ . Further, these operators exert an influence on several equations of electromagnetic and gravitational fields in the complex-octonion curved space directly.

\section{Approximation}

In the curved space $\mathbb{O}_U$ based on the composite radius vector, an approximate theory will be degenerated from the preceding discussions associated with the curved space $\mathbb{O}_U$ , under certain weak-field approximate circumstances. Some inferences of this approximate theory correlate with that of the GR, especially in the case that the energy-momentum tensor is equal to zero.

For the sake of achieving the approximate theory of the curved space based on the composite radius vector, it is necessary to meet the demand of two approximate conditions. a) The bending degree of composite space is quite tiny, and even it is flat, in order to allow separating the arguments of curved space into the spatial parameter and physical quantity. b) The coupling term between the spatial parameter and physical quantity is zero approximately, that is, they could be decoupled.

In the curved composite-space described with the complex octonions, the tangent-frame octonion, $\textbf{\emph{E}}_\alpha = \partial ( \mathbb{H} + k_{rx} \mathbb{X} ) / \partial U^\alpha $ , can be separated into two components, that is, $\textbf{\emph{E}}_\alpha = \textbf{\emph{E}}_{\alpha (H)} + \textbf{\emph{E}}_{\alpha (X)} $ . Herein the component, $\textbf{\emph{E}}_{\alpha (H)} = \partial \mathbb{H} / \partial U^\alpha $ , is related with the spatial parameter, while the component, $\textbf{\emph{E}}_{\alpha (X)} = \partial ( k_{rx} \mathbb{X} ) / \partial U^\alpha$ , is associated with the physical quantity.

Therefore, the metric coefficient in Eq.(14) can be expanded in a Taylor series,
\begin{eqnarray}
g_{\overline{\alpha} \beta (H,X)} && = ( \textbf{\emph{E}}_{\alpha (H)} + \textbf{\emph{E}}_{\alpha (X)} )^\ast \circ ( \textbf{\emph{E}}_{\beta (H)} + \textbf{\emph{E}}_{\beta (X)} )
\nonumber
\\
&& \approx 1 + A_{\overline{\alpha} \beta}    ~~ ,
\end{eqnarray}
where $A_{\overline{\alpha} \beta} = \textbf{\emph{E}}_{\alpha (H)}^\ast \circ \textbf{\emph{E}}_{\beta (X)} + \textbf{\emph{E}}_{\alpha (X)}^\ast \circ \textbf{\emph{E}}_{\beta (H)} $ .
The norm of the component of tangent frame associated with the spatial parameter is close to 1. That is, $\parallel \textbf{\emph{E}}_{\alpha (H)} \parallel \approx 1$ , and $\parallel \textbf{\emph{E}}_{\beta (H)} \parallel \approx 1$ . $ \textbf{\emph{E}}_{\alpha (H)}^\ast \circ  \textbf{\emph{E}}_{\beta (H)} \approx 1 $ .

According to the definition of field potential, the term $( A_{\overline{\alpha} \beta} / k_{rx} )$ possesses the dimension of field potential. But this term is not the field potential, and is merely considered as a term to be equivalent to the field potential. In other words, there are certain complicated relations between the term $( A_{\overline{\alpha} \beta} / k_{rx} )$ with the field potential.

Subsequently, starting from the above metric coefficient, we can deduce the connection coefficient, curvature tensor, and geodetic line and so forth, under the weak-field approximate conditions. Further it is capable of inferring several conclusions in the curved space $\mathbb{O}_U$ , which are consistent with that in the GR, especially in a few vacuum solutions, in case the energy-momentum tensor is zero. What a pity it is that this approximate method, in the curved space $\mathbb{O}_U$ , can merely explain a small quantity of physical phenomena. On the contrary, if we contemplate the influence of these arguments on the divergence, gradient, and curl in the curved space $\mathbb{O}$ , under the weak-field approximate conditions, it will be find that it is possible to explain plenty of physical phenomena in the curved space $\mathbb{O}$ , making use of some equations of electromagnetic and gravitational fields in the curved space $\mathbb{O}$ .

In the GR, the metric coefficient is, $g_{ij} = 1 + h_{ij}$ . And it claimed that the term $h_{ij}$ is equivalent to the gravitational potential in the Newtonian mechanics. In the paper, the minor term $A_{\overline{\alpha} \beta}$ corresponds to the minor term $h_{ij}$ . In other words, the viewpoint of $A_{\overline{\alpha} \beta}$ in the composite space happens to coincide with that of $h_{ij}$ in the GR. As a result, some inferences in the composite space can be reduced into that in the GR. In terms of the analysis of the preceding studies, it will be found that there are a few discrepancies between the paper and GR. a) Metric coefficient. The conjugate applied to the metric coefficient $g_{\overline{\alpha} \beta}$ is the octonion conjugate, rather than the complex octonion. Consequently the scalar term $g_{\overline{\alpha} \beta}$ can be reduced into the term $g_{ij}$ in the GR, under the weak-field approximate conditions. b) Curvature tensor. The GR assumes one energy-momentum tensor, which is still hard to be applied availably up to now. In the composite space, it is supposed that the relation between the spatial parameter and physical quantity can be decoupled under some special circumstances, inferring the relations among the ingredients of curvature tensors. Obviously the physical quantity $R_{\beta \overline{\alpha} \nu ~~(X)}^{~~~~~~\gamma}$ in the paper corresponds to the energy-momentum tensor in the GR. c) Application range. The GR can merely be applied to the gravitational field to explain a few physical phenomena. By contrast, the study in the paper can be applied to the electromagnetic and gravitational fields, making clear several physical phenomena in the curved space $\mathbb{O}_U$ and many physical phenomena in the curved space $\mathbb{O}$ .

On the basis of the above analysis, one can conclude that the complex-octonion composite space is compatible with the GR, and even both of them give mutual support to each other to a certain extent. In most of cases, there is, $\mathbb{H} \gg k_{rx} \mathbb{X}$ , that is, $( \mathbb{H} + k_{rx} \mathbb{X} ) \approx \mathbb{H}$ . As a result, it is not easy to directly sense the influence of term $k_{rx} \mathbb{X}$ , under normal circumstances. Moreover, the paper bears some resemblances to the GR. Under the weak field approximate conditions, the metric coefficients of two theories are similar to each other, and both can be expanded in a Taylor series. Next, the physical meanings of some equations in the two theories are quite close to each other, and both of them claim the standpoint of `the field dominates the space'.

By all appearances, the weak field approximate method in the curved composite-space, $\mathbb{O}_U$ , can be extended into the curved angular-momentum space, $\mathbb{O}_L$ . In the latter, under the weak field conditions, the metric coefficient, $g_{\overline{\alpha} \beta (P,H,X)} = \textbf{\emph{E}}_{\alpha (P,H,X)}^\ast \circ \textbf{\emph{E}}_{\beta (P,H,X)}$ , in Eq.(28) is able to be expanded in a Taylor series as well, achieving the conclusions in accordance with Eq.(42). Finally, it is capable of researching the connection coefficient and curvature tensor of the curved angular-momentum space under the weak field conditions, and their contributions to the divergence, gradient, and curl and other operators, and their influences on some equations of electromagnetic and gravitational fields in the curved space $\mathbb{O}$ .

\section{Conclusions}

According to the Cartesian point of view , the space is only the extension of substance, and does not claim existence on its own. Subsequently, M. Faraday \emph{et al.} introduced the concept of field, while A. Einstein \emph{et al.} found the concept of space-time. Nowadays, the Cartesian point of view is expanded to another one, that is, the space (including the four space-time) is only the extension of field, and does not claim existence on its own. The GR elaborated on the expanded viewpoint, while the paper attempts to explicate the expanded viewpoint as well.

In the curved composite-space $\mathbb{O}_U$ described with the complex octonions, the partial derivative of composite radius vector is utilized for the component of tangent frame. According to the definition of octonion norm, it is able to write out the metric coefficient. From the metric tensor, one can deduce the connection coefficient and curvature tensor of the affine connection space. Analyzing some formulae among ingredients of the curvature tensor reveals that several physical quantities will impact the spatial parameters of curved space. Next, these spatial parameters partly make a contribution to some operators in the curved space $\mathbb{O}$ , exerting an influence on a few equations of electromagnetic and gravitational fields in the curved space $\mathbb{O}$ .

In the curved angular-momentum space $\mathbb{O}_L$ described with the complex octonions, the partial derivative of complex-octonion angular momentum is used for the component of tangent frame. The following researches are similar to that in the curved composite-space $\mathbb{O}_U$ . Also the spatial parameters of the curved angular-momentum space will have an influence on several operators in the curved space $\mathbb{O}$ , impacting partly some equations of electromagnetic and gravitational fields in the curved space $\mathbb{O}$ from a different perspective.

It should be noted the paper discussed only some simple cases about the relations among spatial parameters and physical quantities in the curved composite/angular-momentum space. However it clearly states that the physical quantities will dominate the spatial parameters in the two curved spaces. Later, the spatial parameters will make a contribution towards the divergence, gradient, and curl and other operators, exerting an influence on some equations of electromagnetic and gravitational fields. Under the weak-field approximate conditions, the paper can deduce a few inferences in accordance with that derived from the GR. In the following study, it is going to apply the curved composite/angular-momentum space, described with the complex octonions, to explore the influence of comparatively strong gravitational and electromagnetic fields on the geodetic line and so forth. And it may intend to survey the discrepancies among some tangent spaces of different curved spaces. Moreover, it will contemplate several curved spaces based on other arguments, and their influences on some operators and equations of gravitational and electromagnetic fields.

\appendix

\section{Connection coefficient}

In the curved space described with the complex octonions, there is a relation between the metric coefficient and connection coefficient. The definition of metric coefficient must meet the demand of this relation, deducing the connection coefficient from the metric coefficient.

Multiplying the component $\textbf{\emph{e}}_\lambda^\ast$ from the left of the definition,
\begin{equation}
\partial^2 \mathbb{H} / \partial u^\beta \partial u^\gamma = \Gamma_{\beta \gamma }^\alpha \textbf{\emph{e}}_\alpha    ~   ,
\end{equation}
yields,
\begin{equation}
( \partial \mathbb{H}^\ast / \partial u^\lambda ) \circ ( \partial^2 \mathbb{H} / \partial u^\beta \partial u^\gamma ) = g_{\overline{\lambda} \alpha} \Gamma_{\beta \gamma }^\alpha   ~   ,
\end{equation}
while multiplying the component $\textbf{\emph{e}}_\lambda$ from the right of the conjugate of Eq.(43) produces,
\begin{equation}
( \partial^2 \mathbb{H}^\ast / \partial u^\beta \partial u^\gamma ) \circ ( \partial \mathbb{H} / \partial u^\lambda ) = \overline{\Gamma_{\beta \gamma }^\alpha}  g_{\overline{\alpha} \lambda} ~   ,
\end{equation}
where $\partial^2 \mathbb{H}^\ast / \partial u^\beta \partial u^\gamma = \overline{\Gamma_{\beta \gamma }^\alpha} \textbf{\emph{e}}_\alpha^\ast $ . $\Gamma_{\beta \gamma }^\alpha$ and $\overline{\Gamma_{\beta \gamma }^\alpha}$ are coefficients.

From the last two equations, the partial derivative of the metric tensor, $g_{\overline{\lambda} \gamma} = \textbf{\emph{e}}_\lambda^\ast \circ \textbf{\emph{e}}_\gamma $ , with respect to the coordinate value $u^\beta$ generates,
\begin{equation}
g_{\overline{\lambda} \alpha} \Gamma_{\beta \gamma }^\alpha + \overline{\Gamma_{\beta \lambda }^\alpha}  g_{\overline{\alpha} \gamma} = \partial g_{\overline{\lambda} \gamma} / \partial u^\beta  ~ ,
\end{equation}
similarly there are,
\begin{eqnarray}
g_{\overline{\beta} \alpha} \Gamma_{\gamma \lambda }^\alpha + \overline{\Gamma_{\gamma \beta }^\alpha}  g_{\overline{\alpha} \lambda} = \partial g_{\overline{\beta} \lambda} / \partial u^\gamma  ~ ,
\\
g_{\overline{\gamma} \alpha} \Gamma_{\lambda \beta }^\alpha + \overline{\Gamma_{\lambda \gamma }^\alpha}  g_{\overline{\alpha} \beta} = \partial g_{\overline{\gamma} \beta} / \partial u^\lambda  ~ .
\end{eqnarray}

From the last three equations, there are,
\begin{equation}
\Gamma_{ \overline{\lambda} , \beta \gamma } = (1/2) ( \partial g_{ \overline{\gamma} \lambda } / \partial u^\beta + \partial g_{ \overline{\lambda} \beta } / \partial u^\gamma - \partial g_{ \overline{\gamma} \beta } / \partial u^\lambda )    ~  ,
\end{equation}
where $ [ ( g_{\overline{\lambda} \alpha} \Gamma_{\beta \gamma }^\alpha )^\ast ]^T = \overline{\Gamma_{\gamma \beta }^\alpha}  g_{\overline{\alpha} \lambda} $ ,
and $ [ ( \overline{\Gamma_{\gamma \beta }^\alpha}  g_{\overline{\alpha} \lambda} )^\ast ]^T = g_{\overline{\lambda} \alpha} \Gamma_{\beta \gamma }^\alpha $ .

On the other hand, multiplying the component $\textbf{\emph{e}}_\lambda^\ast$ from the left of the definition,
\begin{equation}
\partial^2 \mathbb{H} / \partial \overline{u^\beta} \partial u^\gamma = \Gamma_{\overline{\beta} \gamma }^\alpha \textbf{\emph{e}}_\alpha    ~   ,
\end{equation}
deduces,
\begin{equation}
( \partial \mathbb{H}^\ast / \partial u^\lambda ) \circ ( \partial^2 \mathbb{H} / \partial \overline{u^\beta} \partial u^\gamma ) = g_{\overline{\lambda} \alpha} \Gamma_{\overline{\beta} \gamma }^\alpha   ~   ,
\end{equation}
while multiplying the component $\textbf{\emph{e}}_\lambda$ from the right of the conjugate of Eq.(50) produces,
\begin{equation}
( \partial^2 \mathbb{H}^\ast / \partial \overline{u^\beta} \partial u^\gamma ) \circ ( \partial \mathbb{H} / \partial u^\lambda ) = \overline{\Gamma_{\overline{\beta} \gamma }^\alpha}  g_{\overline{\alpha} \lambda} ~   ,
\end{equation}
where $\partial^2 \mathbb{H}^\ast / \partial \overline{u^\beta} \partial u^\gamma = \overline{\Gamma_{\overline{\beta} \gamma }^\alpha} \textbf{\emph{e}}_\alpha^\ast $ . $\Gamma_{\overline{\beta} \gamma }^\alpha$ and $\overline{\Gamma_{\overline{\beta} \gamma }^\alpha}$ are coefficients.

By means of the last two equations, the partial derivative of the metric tensor, $g_{\overline{\lambda} \gamma} = \textbf{\emph{e}}_\lambda^\ast \circ \textbf{\emph{e}}_\gamma $ , with respect to the coordinate value $\overline{u^\beta}$ generates,
\begin{equation}
g_{\overline{\lambda} \alpha} \Gamma_{\overline{\beta} \gamma }^\alpha + \overline{\Gamma_{\overline{\beta} \lambda }^\alpha}  g_{\overline{\alpha} \gamma} = \partial g_{\overline{\lambda} \gamma} / \partial \overline{u^\beta}  ~ ,
\end{equation}
similarly there are,
\begin{eqnarray}
g_{\overline{\beta} \alpha} \Gamma_{\overline{\gamma} \lambda }^\alpha + \overline{\Gamma_{\overline{\gamma} \beta }^\alpha}  g_{\overline{\alpha} \lambda} = \partial g_{\overline{\beta} \lambda} / \partial \overline{u^\gamma}  ~ ,
\\
g_{\overline{\gamma} \alpha} \Gamma_{\overline{\lambda} \beta }^\alpha + \overline{\Gamma_{\overline{\lambda} \gamma }^\alpha}  g_{\overline{\alpha} \beta} = \partial g_{\overline{\gamma} \beta} / \partial \overline{u^\lambda}  ~ .
\end{eqnarray}

From the last three equations, there exist,
\begin{equation}
\Gamma_{ \overline{\lambda} , \overline{\beta} \gamma } = (1/2) ( \partial g_{ \overline{\gamma} \lambda } / \partial \overline{u^\beta} + \partial g_{ \overline{\lambda} \beta } / \partial \overline{u^\gamma} - \partial g_{ \overline{\gamma} \beta } / \partial \overline{u^\lambda} )    ~  ,
\end{equation}
where $ [ ( g_{\overline{\lambda} \alpha} \Gamma_{\overline{\beta} \gamma }^\alpha )^\ast ]^T = \overline{\Gamma_{\overline{\gamma} \beta }^\alpha}  g_{\overline{\alpha} \lambda} $ , and $ [ ( \overline{\Gamma_{\overline{\gamma} \beta }^\alpha}  g_{\overline{\alpha} \lambda} )^\ast ]^T = g_{\overline{\lambda} \alpha} \Gamma_{\overline{\beta} \gamma }^\alpha $ .

\section{Curvature tensor}

In the curved space described with the complex octonions, for a tensor $Y^\gamma$ with contravariant rank 1, the covariant derivative is,
\begin{equation}
\nabla_\beta Y^\gamma = \partial Y^\gamma / \partial u^\beta + \Gamma _{\lambda \beta}^\gamma Y^\lambda ~ , ~~~~~
\nabla_{\overline{\alpha}} Y^\gamma = \partial Y^\gamma / \partial \overline{u^\alpha} + \Gamma _{\lambda \overline{\alpha}}^\gamma Y^\lambda  ~,
\end{equation}
meanwhile, for a mixed tensor, $Z_\nu^\gamma$ , with contravariant rank 1 and covariant rank 1, the covariant derivative can be written as,
\begin{eqnarray}
\nabla_\beta Z_\nu^\gamma = \partial Z_\nu^\gamma / \partial u^\beta - \Gamma_{\beta \nu}^\lambda Z_\lambda^\gamma + \Gamma_{\beta \lambda}^\gamma Z_\nu^\lambda  ~ ,
\\
\nabla_{\overline{\alpha}} Z_\nu^\gamma = \partial Z_\nu^\gamma / \partial \overline{u^\alpha} - \Gamma_{\overline{\alpha} \nu}^\lambda Z_\lambda^\gamma + \Gamma_{\overline{\alpha} \lambda}^\gamma Z_\nu^\lambda  ~ ,
\end{eqnarray}
where $Y^\gamma$ and $Z_\nu^\gamma$ both are scalar.

Apparently, the above equations deduce,
\begin{equation}
\nabla_{\overline{\alpha}} ( \nabla_\beta Y^\gamma ) = \partial ( \nabla_\beta Y^\gamma ) / \partial \overline{u^\alpha} - \Gamma_{\beta \overline{\alpha} }^\lambda ( \nabla_\lambda Y^\gamma )
+ \Gamma_{\lambda \overline{\alpha} }^\gamma ( \nabla_\beta Y^\lambda )   ~ ,
\end{equation}
similarly there is,
\begin{equation}
\nabla_\beta ( \nabla_{\overline{\alpha}} Y^\gamma ) = \partial ( \nabla_{\overline{\alpha}} Y^\gamma ) / \partial u^\beta - \Gamma_{\overline{\alpha} \beta }^\lambda ( \nabla_\lambda Y^\gamma )
+ \Gamma_{\lambda \beta }^\gamma ( \nabla_{\overline{\alpha}} Y^\lambda )   ~ .
\end{equation}

As a result, making a subtraction in last two equations yields,
\begin{eqnarray}
&& \nabla_{\overline{\alpha}} ( \nabla_\beta Y^\gamma )  -  \nabla_\beta ( \nabla_{\overline{\alpha}} Y^\gamma )
\nonumber
\\
= && \{ \partial \Gamma_{\nu \beta}^\gamma / \partial \overline{u^\alpha} + \Gamma_{\lambda \overline{\alpha}}^\gamma \Gamma_{\nu \beta}^\lambda
- \partial \Gamma_{\nu \overline{\alpha}}^\gamma /\partial u^\beta - \Gamma_{\lambda \beta}^\gamma \Gamma_{\nu \overline{\alpha}}^\lambda \} Y^\nu
+ T_{\beta \overline{\alpha}}^\lambda (\nabla_\lambda Y^\gamma)  ~  ,
\end{eqnarray}
where $ \partial ( \partial Y^\gamma / \partial \overline{u^\alpha} ) / \partial u^\beta - \partial ( \partial Y^\gamma / \partial u^\beta ) / \partial \overline{u^\alpha} = 0 $ . The torsion tensor is, $ T_{\beta \overline{\alpha}}^\lambda = \Gamma_{\overline{\alpha} \beta }^\lambda - \Gamma_{\beta \overline{\alpha}}^\lambda $ .

In the case there is, $ T_{\beta \overline{\alpha}}^\lambda = 0 $ , the above can be reduced to,
\begin{eqnarray}
\nabla_{\overline{\alpha}} ( \nabla_\beta Y^\gamma )  -  \nabla_\beta ( \nabla_{\overline{\alpha}} Y^\gamma ) = R_{\beta \overline{\alpha} \nu}^{~~~~~~\gamma} Y^\nu  ~  ,
\end{eqnarray}
where the curvature tensor is, $ R_{\beta \overline{\alpha} \nu}^{~~~~~~\gamma} = \partial \Gamma_{\nu \beta}^\gamma / \partial \overline{u^\alpha} - \partial \Gamma_{\nu \overline{\alpha}}^\gamma /\partial u^\beta + \Gamma_{\lambda \overline{\alpha}}^\gamma \Gamma_{\nu \beta}^\lambda - \Gamma_{\lambda \beta}^\gamma \Gamma_{\nu \overline{\alpha}}^\lambda $ .

\section{Tangent space}

In the curved spaces described with the complex octonions, there are some particular situations, in which the tangent spaces are different from the underlying spaces. They include but not limited to the following cases:

1) The underlying space is the complex octonion space $\mathbb{O}$ , while some tangent spaces respectively relate with the complex-octonion radius vector $\mathbb{H}$ , composite radius vector $\mathbb{U}$ , and angular momentum $\mathbb{L}$ and so forth. Therefore their components of tangent frames are respectively the partial derivatives of the complex-octonion radius vector, composite radius vector, and angular momentum and so on, with respect to the coordinate value $u^\alpha$ of complex-octonion radius vector.

2) The underlying space is the complex-octonion composite space $\mathbb{O}_U$ , while several tangent spaces are respectively relevant to the complex-octonion composite radius vector, angular momentum, and torque $\mathbb{W}$ and so forth. Accordingly their components of tangent frames are respectively the partial derivatives of the complex-octonion composite radius vector, angular momentum, and torque and so on, with respect to the coordinate value $U^\alpha$ of complex-octonion composite radius vector.

3) The underlying space is the complex-octonion angular momentum space $\mathbb{O}_L$ , while a part of tangent spaces are respectively involved with the complex-octonion angular momentum, torque, force $\mathbb{N}$ and so forth. Consequently their components of tangent frames are respectively the partial derivatives of the complex-octonion angular momentum, torque, and force and so on, with respect to the component $L^\alpha$ of complex-octonion angular momentum.

Apparently, in these curved spaces, there are some special cases, that is, the metric coefficient is a dimensionless parameter, and able to be approximately degenerated into, $g_{\overline{\alpha} \beta} \approx 1 + h_{\alpha \beta} $ . For example, a) The underlying space and tangent space both are the complex octonion space $\mathbb{O}$ . b) The underlying space and tangent space both are the complex-octonion composite space $\mathbb{O}_U$ . c) The underlying space and tangent space both are the complex-octonion angular momentum space $\mathbb{O}_L$ . d) The underlying space is the complex octonion space $\mathbb{O}$ , while the tangent space is the complex-octonion composite space $\mathbb{O}_U$ .

For the majority of the curved spaces described with the complex octonions, it may cause the metric coefficient to be seized of one certain dimension in the physics. As a result, they are incapable of deducing some equations to meet the requirement of the academic thought, that is, the existence of field will dominate the bending of space. The weak approximate method may be not suitable to these curved spaces, and even it may be hard to comprehend a few physical meanings may be in these curved spaces.

\begin{acknowledgements}
The author is indebted to the anonymous referees for their constructive comments on the previous manuscript. This project was supported partially by the National Natural Science Foundation of China under grant number 60677039.
\end{acknowledgements}



\end{document}